\documentclass[%
reprint,
showpacs,
showkeys,
nofootinbib,
nobibnotes,
amsmath,
amssymb,
aps, 
superscriptaddress
]{revtex4-1}

\usepackage[english]{babel}
\usepackage[utf8x]{inputenc}
\usepackage[T1]{fontenc}
\usepackage{lineno}
\usepackage{caption}
\usepackage[skip=0cm,list=true,labelfont=it]{subcaption}
\usepackage{amsmath}
\usepackage{graphicx}
\usepackage[colorinlistoftodos]{todonotes}
\usepackage[colorlinks=true, allcolors=blue]{hyperref}
\usepackage{epstopdf}
\usepackage{amssymb}
\usepackage{color}
\usepackage{url}
\usepackage[justification=justified]{subcaption}
\usepackage[justification=justified, figurename=FIG.]{caption}
\usepackage{float}

\restylefloat*{figure}

\usepackage{dcolumn}
\usepackage{bm} 
\usepackage{hyperref} 

\usepackage{siunitx}
\usepackage{textcomp} 
\usepackage{multirow}
\usepackage{tabularx} 
\usepackage{ragged2e}

\newcommand{\Lagr}{\mathcal{L}}

\begin{document}


\title{Projected sensitivities of the LUX-ZEPLIN (LZ) experiment to new physics via low-energy electron recoils}{

\author{D.S.~Akerib}
\affiliation{SLAC National Accelerator Laboratory, Menlo Park, CA 94025-7015, USA}
\affiliation{Kavli Institute for Particle Astrophysics and Cosmology, Stanford University, Stanford, CA  94305-4085 USA}

\author{A.K.~Al Musalhi}
\affiliation{University of Oxford, Department of Physics, Oxford OX1 3RH, UK}

\author{S.K.~Alsum}
\affiliation{University of Wisconsin-Madison, Department of Physics, Madison, WI 53706-1390, USA}

\author{C.S.~Amarasinghe}
\affiliation{University of Michigan, Randall Laboratory of Physics, Ann Arbor, MI 48109-1040, USA}

\author{A.~Ames}
\affiliation{SLAC National Accelerator Laboratory, Menlo Park, CA 94025-7015, USA}
\affiliation{Kavli Institute for Particle Astrophysics and Cosmology, Stanford University, Stanford, CA  94305-4085 USA}

\author{T.J.~Anderson}
\affiliation{SLAC National Accelerator Laboratory, Menlo Park, CA 94025-7015, USA}
\affiliation{Kavli Institute for Particle Astrophysics and Cosmology, Stanford University, Stanford, CA  94305-4085 USA}

\author{N.~Angelides}
\affiliation{University College London (UCL), Department of Physics and Astronomy, London WC1E 6BT, UK}

\author{H.M.~Ara\'{u}jo}
\affiliation{Imperial College London, Physics Department, Blackett Laboratory, London SW7 2AZ, UK}

\author{J.E.~Armstrong}
\affiliation{University of Maryland, Department of Physics, College Park, MD 20742-4111, USA}

\author{M.~Arthurs}
\affiliation{University of Michigan, Randall Laboratory of Physics, Ann Arbor, MI 48109-1040, USA}

% ...


\begin{abstract}
\noindent LUX-ZEPLIN (LZ) is a dark matter detector expected to obtain world-leading sensitivity to weakly interacting massive particles (WIMPs) interacting via nuclear recoils with a $\sim$7-tonne xenon target mass. This manuscript presents sensitivity projections to several low-energy signals of the complementary \emph{electron} recoil signal type:  1)  an effective neutrino magnetic moment and 2) an effective neutrino millicharge, both for pp-chain solar neutrinos, 3) an axion flux generated by the Sun, 4) axion-like particles forming the galactic dark matter, 5) hidden photons, 6) mirror dark matter, and 7) leptophilic dark matter.  World-leading sensitivities are expected in each case, a result of the large 5.6~t 1000~d exposure and low expected rate of electron recoil backgrounds in the $<$100~keV energy regime.  A consistent signal generation, background model and profile-likelihood analysis framework is used throughout. 

\end{abstract}


\maketitle

\section{Introduction}
\noindent

LUX-ZEPLIN (LZ) is a liquid xenon (LXe) time projection chamber (TPC) currently being commissioned at the Sanford Underground Research Facility (SURF) in the US~\cite{surf}.  It is the latest in a series of increasingly large LXe TPCs optimized for sensitivity to rare keV-scale nuclear recoil signatures indicative of weakly interacting massive particle (WIMP)
dark matter~\cite{zeplini,
zeplinii_1,
zepliniii-ssr, 
LUX_complete,
X10_SI,
X100_477d_SI,
X1T_1tyear,
PX1,
PX2_99d}. 
These detectors achieve extremely low background rates thanks to the very high purity of the LXe material itself and the self-shielding of the LXe, which inhibits the penetration of external radiation into the central fiducial LXe volume.

While the design goal of these dark matter detectors is sensitivity to nuclear recoils, several aspects of that optimization (a keV-scale energy threshold, large size, and low background) together provide complementary sensitivity to low-energy, low-rate processes of electron recoil (ER) type.  In this paper we present sensitivity studies for several novel processes that would produce low-rate ER signatures with energy depositions smaller than 100~keV. Section~II describes the essential aspects of the LZ instrument.  Section~III gives a brief description of the analysis framework used. Section IV presents the background model with particular emphasis on low-energy ER backgrounds. In Section~V, the signal models for each of the different novel physics processes are presented.  Section~VI briefly presents the profile likelihood ratio (PLR) statistical method used to determine the sensitivity reach of LZ to each of the possible signals.  Results are presented in Section~VII.  For each signal model, a projection for 90\% confidence level (CL) exclusion sensitivity is presented.  For selected signal models, projected sensitivity for 3$\sigma$ evidence is also presented. Conclusions are presented in Section~VIII.

\section{Description of the experiment}
\noindent 
The LZ experiment is located at a depth of 1480~m (4300~m water equivalent), which reduces the cosmic ray muon flux by a factor of~$\sim$10$^{6}$~\cite{LZ_simulations}. The LXe detector is surrounded by an instrumented water tank that shields against backgrounds, including gamma rays originating in the surrounding rock.  The LXe cryostat is immediately surrounded by an active veto layer of Gd-loaded liquid scintillator (the `outer detector'), primarily for tagging nuclear recoils in the TPC, but also effective at further reducing ER backgrounds.  The LXe is contained within a double-walled cylindrical vessel fabricated from low background titanium~\cite{titanium}.  Surrounding the TPC on the sides and bottom is a $\sim$2-tonne layer of optically-isolated LXe  (the `LXe skin') that is separately instrumented and serves as an active veto to further reduce backgrounds in the central volume.  The central TPC volume includes a liquid/gas interface near the top of the volume, such that a thin layer of gaseous xenon is maintained for signal generation purposes. A vertical electric field, anticipated to be $\sim$310~V/cm in the LXe, is applied using an anode placed within the gas layer and a cathode towards the bottom of the liquid. Incident radiation scattering within the LXe generates electron and/or nuclear recoils, in turn leading to electronic excitations and ionizations. De-excitations produce primary scintillation (denoted `S1'), while ionization electrons that escape recombination are drifted by the electric field to the liquid surface. A `gate' grid is located just under the liquid surface to assist with extraction of these electrons to the gas phase, from where a delayed second signal (`S2') of electroluminescence is generated. The electroluminescence gain in the gas region allows observation of single electrons emitted from the liquid surface.  In analyses requiring an S1 signal (as is the case in this manuscript), it is the S1 signal that sets the energy threshold.  The simulations of this work project the 10\%, 50\%, and 90\% detection efficiency thresholds to be 1.04, 1.45, and 2.14~keV respectively.

Arrays of high quantum efficiency, vacuum ultraviolet sensitive, low radioactivity photomultiplier tubes (PMTs), Hamamatsu model R11410-20, are placed both above (253~PMTs) and below (241~PMTs) the TPC volume. The S1 signal is relatively small, while the delayed secondary S2 is larger, with the signal amplitude ratio S2/S1 typically larger for ER interactions and smaller for NR interactions. The horizontal location of an energy deposition may be reconstructed from the distribution of S2 signal amplitudes in the upper PMT array, and the vertical location can be inferred from the time delay between the S1 and S2. A large fraction of energy depositions originate from trace radioactivity of the surrounding materials, and thus are observed near the edges of the LXe volume. The precise event position information given by the TPC configuration enables the definition in offline analysis of a central `fiducial' volume of extremely low background. The LZ TPC surrounds a LXe mass of 7.0~tonnes; the central fiducial volume is taken as 5.6~tonnes. The detector's low-energy ER response will be characterized using radioactive sources injected into the fiducial volume, including the beta decaying sources $^{3}$H and $^{14}$C (in the form of $^{3}$H-labeled and $^{14}$C-labeled CH$_4$), and the monoenergetic sources $^{83\rm{m}}$Kr (41~keV) and $^{131\rm{m}}$Xe (164~keV).  External gamma ray sources and neutron sources will also be used.  Further details of the LZ apparatus may be found in Refs.~\cite{LZ_TDR} and~\cite{LZdetector}.

Previous papers have described expected LZ sensitivities to signatures of new physics:  WIMP sensitivity is described in~\cite{LZ_wimp} and~\cite{newDPEpaper}, and sensitivity to the neutrinoless double beta decay of $^{136}$Xe is described in~\cite{LZ_NDBD}.  The background models in those studies and this paper are built by first conducting a campaign of material assays~\cite{LZ_screening} and then simulating the effects of those trace amounts of radioactivity in the LZ geometry~\cite{LZ_simulations}.  Also simulated is the production and collection of S1 and S2 light signals.  The observed signals are measured in units of photons detected (`phd'), an observable that accounts for the occurrence of double photoelectron emission from PMTs~\cite{Faham_2015, PhysRevD.97.102008}. The average fraction of S1 scintillation photons that are detected is denoted $g_1$ and is predicted by simulation to be 0.119~phd/photon, while the average number of photoelectrons generated by an ionization electron extracted from the liquid surface is denoted $g_2$ and is predicted to be~79.2~phd/electron. While single photoelectrons produce a signal clearly separable from electronic noise, an anticipated kHz-scale dark rate of single electron emission (within the PMTs) results in a 3-fold S1 coincidence requirement, meaning a 3~phd S1 threshold.  (This coincidence requirement can be reduced as in~\cite{LUX_singlephd} and~\cite{newDPEpaper}, but this technique is not taken advantage of by this analysis.)  All detector response characteristics in this analysis match previous LZ sensitivity papers, see Refs.~\cite{LZ_wimp,LZ_NDBD}.

\section{Analysis Framework}\label{sec:AnalFrame}

\noindent
The LZ simulation chain~\cite{LZ_simulations} was used to generate the expected distributions of detector observables for radiation emitted from each of the background components and for each  signal model. For those background components which are either non-uniform in spatial distribution or possibly registered as multiple scatters, we begin with an in-house software package that employs \textsc{Geant4}~\cite{Geant4} to track particles as they propagate through the detector. Several of the physics routines have been modified to better model interactions with xenon and gadolinium-loaded scintillator. Energy deposits, timing and spatial information for the interaction are recorded and the photon-level signal response for each PMT generated. The timing and spatial information are used to enable the application of data selection cuts to the simulated data. Scatters are accepted as `single scatter' events if there is no coincident energy deposit in the active skin or outer detector volumes (given nominal thresholds) and if their several energy deposits in the LXe are of small spatial extent.  Quantitatively, this spatial extent requirement is defined as having an energy-weighted standard deviation $<$3~cm in the radial direction and $<$0.2~cm in the vertical direction, imitating the effect of selection cuts based on S2 light distribution and pulse shape.

For those event types which are both uniform in space and always appear as single-scatters, the \textsc{Geant4} framework is not necessary, and we instead use theoretical energy spectra generated analytically, giving the expected number of events per unit mass and time, which can then be scaled according to exposure.  This simpler framework is used for each of the signal models described in Section~\ref{sec:signals}, along with the simplest background components ($^{136}$Xe decay, $^{124}$Xe decay, and both ER and NR neutrino scatters).

After these background and signal energy spectra are compiled, each is then used as input to the Noble Element Simulation Technique (NEST) software~\cite{NEST2011, NESTv2.0}, where ER and NR energy deposits are translated to the detector observables S1 and S2.  The position-dependence of signal collection efficiencies is included in that translation to S1 and S2.  Finally, that position-dependence is removed by applying position-dependent `corrections' as planned to actual data, to form the final analysis quantities S1c and S2c. These simulations of signal production and collection are based on detector response parameters specific to the LZ detector (including for example $g_1$, $g_2$, and electric field amplitude).

After signals are simulated, selections are applied mimicking those selections that will be applied to real data.  Valid events must have an S1 in which at least three PMTs register at least a single photoelectron, while the uncorrected S2 signal is required to be greater than 415~phd, equivalent to $\approx$5 electrons emitted from the liquid surface, ensuring that the radial position of the scatter is adequately reconstructed. Fiducial volume cuts $r < 68.8$~cm from the center and $1.5 < z < 132.1$~cm above the cathode are applied to define the 5.6-tonne fiducial mass, consistent with~\cite{LZ_wimp}.

\section{Backgrounds} \label{sec:backgrounds}

\noindent
LZ sensitivity to rare ER signals is limited by a combination of radiogenic and cosmogenic backgrounds. With limited exceptions, the backgrounds of this manuscript are modeled as in~\cite{LZ_wimp}, and for that reason we do not repeat a detailed discussion here, describing only changes relative to that work.  The first change is a more accurate treatment of the atomic binding effects in the solar neutrino electron scattering spectrum~\cite{Athoy_thesis}. While {\it pp} and $^{7}$Be neutrinos dominate the neutrino scattering rate, there are additional contributions from the CNO, $^{8}$B, {\it hep} and {\it pep} mechanisms. Scattering rates are taken from the relativistic random phase approximation (RRPA)~\cite{jwchen}, used directly up to 30~keV.  Above 30~keV, a simpler method based on the stepping approximation~\cite{1997PAN....60.1859K} is used, scaled to agree with RRPA at 30~keV.

The second change is the inclusion of two-neutrino double electron capture (DEC) of $^{124}$Xe as a background contribution. We include the monoenergetic peaks from the three most frequent K- and L-shell capture combinations: 64.3~keV (KK-shell capture, branching fraction 76.6\%), 36.7~keV (KL-shell capture, 23\%), and 9.8~keV (LL-shell capture, 1.7\%).  The decay rate of (1.8$\pm$0.5$\pm$0.1)$\times$10$^{22}$~years is informed by the XENON1T experiment's recently-reported evidence for KK-shell capture~\cite{XE124DEC}.  Neutron capture on $^{124}$Xe (mostly on the unshielded xenon outside the water tank) will result in constant low-level production of $^{125}$I, with decay rate in the TPC highly dependent on iodine's specific purification timescale.  Conservatively assuming a 10d purification halflife, we anticipate $\sim$100 $^{125}$I decays in the 5.6~t 1000~d exposure.  Because this number is small in comparison to the $^{124}$Xe KK-shell background (2,527 counts expected), and nearly degenerate in energy (67.3~keV), $^{125}$I is not included in the background model of this analysis.

A third change is the introduction of two `unexpected but possible' backgrounds components.  Although this is a broad class, we introduce two such components which have a relevant rate in the few-keV regime where XENON1T recently reported evidence of an excess~\cite{xenon1t_2020,matthew,X1T3H}:  $^{37}$Ar (treated as a monoenergetic peak at 2.8~keV) and $^{3}$H (a beta decay with a broad peak at $\sim$3~keV and endpoint of 18.6~keV).  These two backgrounds are expected to have negligible long-term contribution and are therefore not included in the projections of exclusion sensitivity.  However, if an excess is seen that is consistent with both new physics and such a background, it would be difficult to rule out the background hypothesis and claim discovery.  Both radioisotopes can be formed via spallation, for example, and there may be an unknown production path in detector materials.  Indeed, our own calculations suggest cosmogenically produced $^{37}$Ar will be present at some level in the earliest LZ data~\cite{37Ar_future}.  We therefore conservatively include such unexpected but possible backgrounds when quantifying sensitivity to evidence of signal (the exclusion of the background-only hypothesis).  The method of this inclusion is described in Sec~\ref{sec:stats}.}

Recently, several authors (see Refs.~\cite{scottha_betas, xenon1t_2020}) have offered improved descriptions of the spectral shape of the $^{214}$Pb, $^{212}$Pb, and $^{85}$Kr beta decay spectra, particularly at low energies.  These refinements are not included in the present work, but the effect of such a reduction in the rate of these background components is discussed in Subsection~\ref{sec:variedbetas}.

\begin{figure}
\includegraphics[width=3.25in]{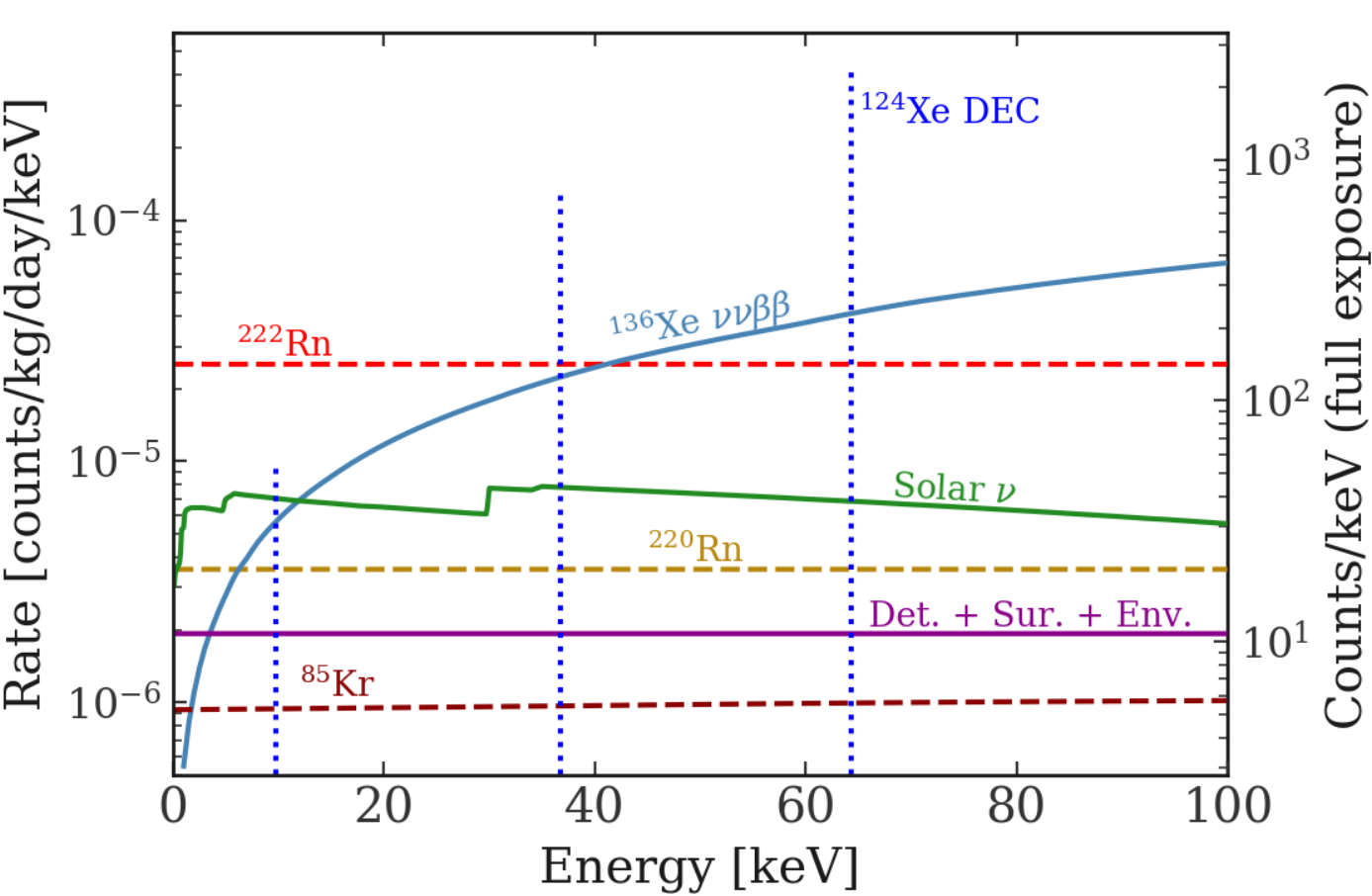}
\caption{The ER backgrounds expected in LZ, after application of veto anti-coincidence, single scatter, and fiducial volume selections. The left axis indicates rate in standard units;  the right axis indicates counts per keV in the anticipated 5.6~t 1000~d LZ exposure.  This is true energy, where effects of detector resolution and threshold are not included.  The three dashed curves indicate the three species that are present as contaminants within the LXe, for which some uncertainty exists on their final concentrations.  The curve labelled `Det.+Sur.+Env.' identifies contributions from the Compton scattering of $\gamma$-rays emitted from the bulk and surfaces of detector components, and from the laboratory and rock environment. Three monoenergetic peaks from $^{124}$Xe double electron capture decay are indicated as lines (for which the units are counts/kg/day and counts on the left and right axes, respectively). \label{fig:bgspec}}
\end{figure}

\begin{figure}
\includegraphics[width=3.25in]{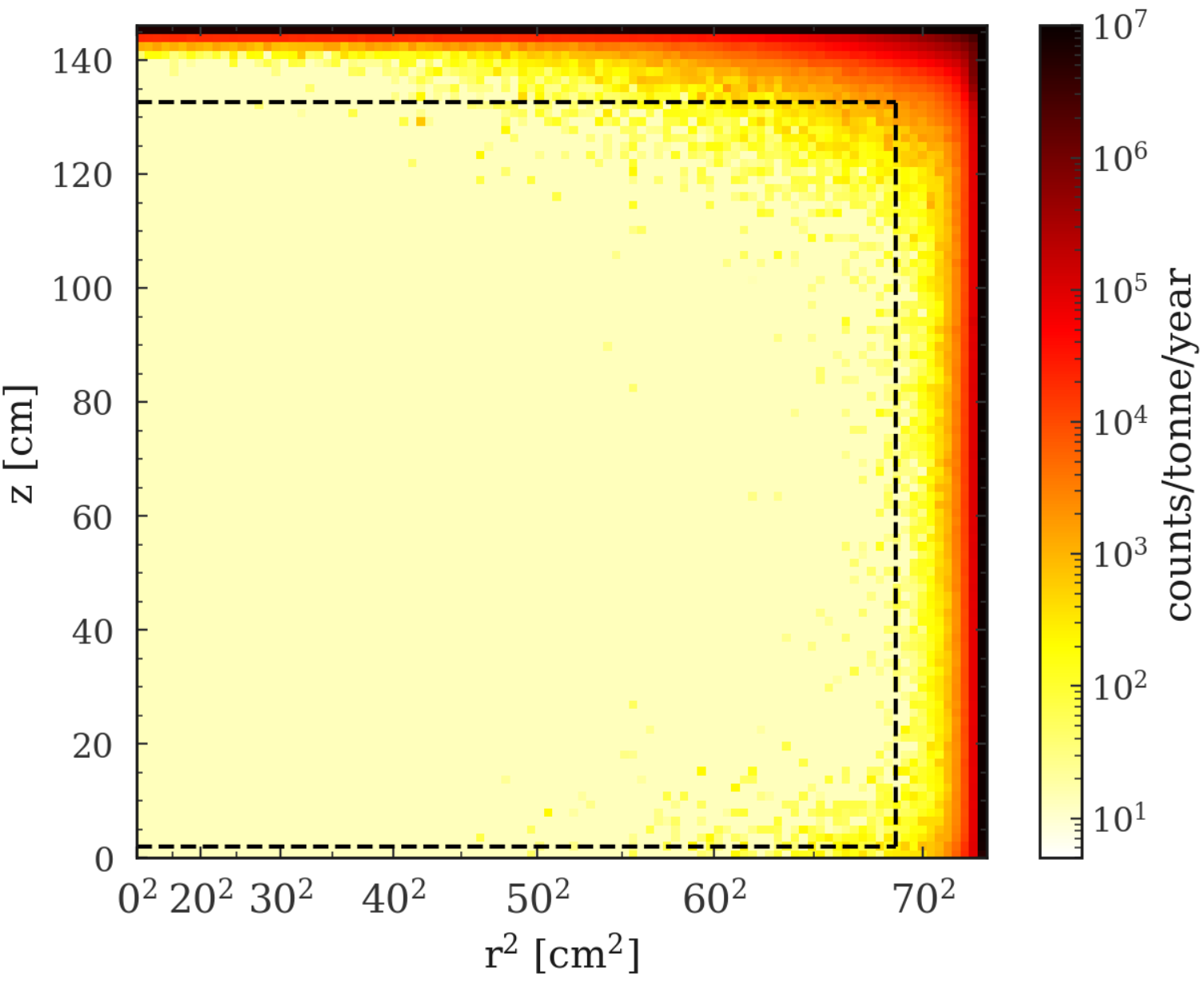}
\caption{\label{fig:bgrz} Spatial distribution of all ER backgrounds passing single scatter and veto anti-coincidence selection criteria, with energy less than 100~keV. The 5.6~tonne fiducial region is indicated by the dashed black line.}
\end{figure}

Figure~\ref{fig:bgspec} shows the predicted ER background event rates, both in terms of a rate (counts/[kg day keV]), and in terms of counts (per keV) for the full LZ exposure of 1000~days.  Here we plot the spectra of true deposited energy, meaning detector resolution and threshold are not included.  NR background rates are dominated by coherent elastic scattering of $^{8}$B neutrinos, and while NR backgrounds are included in the sensitivity projections of this manuscript, they are not shown in Figure~\ref{fig:bgspec} because their effect on ER sensitivities is negligible.  Quantitatively, over the lifetime of LZ, we expect $<$1 NR appearing within the 3$\sigma$ (99.7\%) contours of the ER region of the \{S1c, logS2c\} plane.  The total background rate is dominated below $\sim$40~keV by the decay of $^{222}$Rn progeny (predominantly the ground state beta decay of $^{214}$Pb), and is dominated at higher energies by the two-neutrino double beta decay ($\nu\nu\beta\beta$) of $^{136}$Xe.  All of these background components are uniform in rate throughout the TPC with the exception of the component labeled `Detector + Surface + Environment'.  This component leads to spatial dependence in the total background rate and to the definition of the fiducial volume selection, illustrated in Figure~\ref{fig:bgrz}.  In both Figures~\ref{fig:bgspec} and~\ref{fig:bgrz}, veto and single scatter selections have been applied.

\section{Signal models}\label{sec:signals}
\noindent
In this Section we present seven examples of novel physics processes that would produce low energy ER signals in LZ. For each new physics process, the theoretical motivation is briefly recapped and the expected deposited energy spectra and overall interaction rate are described as a function of the relevant cross section, coupling constant, or other physical parameter.  As described in Section III, these true energy spectra are then translated into the experimental observables of S1c and S2c, before testing their observability given the  expected backgrounds.  The range of models chosen includes a variety of spectral shapes, from mono-energetic signatures to gently sloping spectra.

\begin{figure*}[ht]
\includegraphics[width=\textwidth]{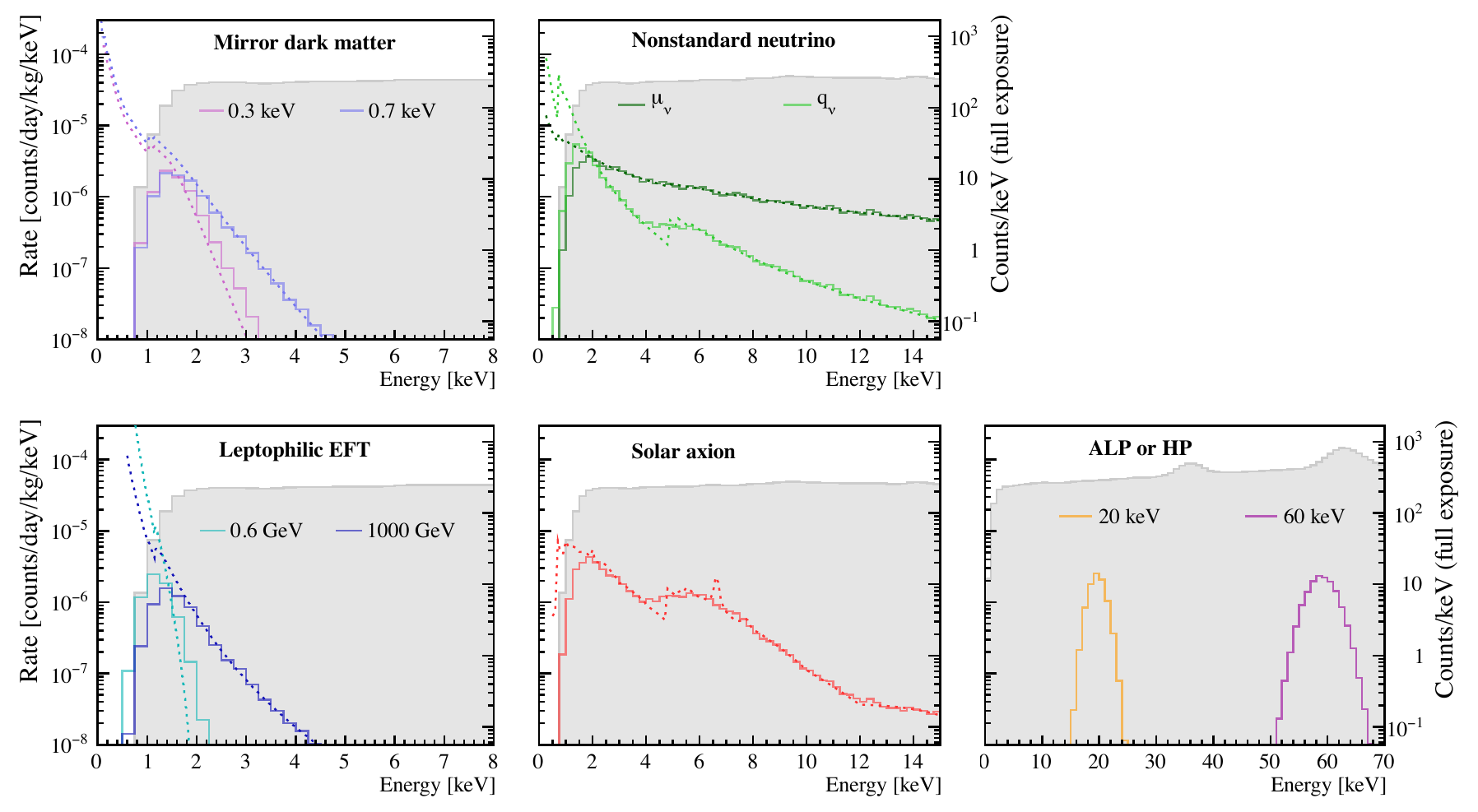}
\caption{\label{fig:allSpec}
ER signal spectra studied in this work.  True deposited energies are shown using a dotted line and reconstructed energies that include smearing and threshold effects with solid lines. The expected background, also including experimental effects, is shown as the grey region.  Each signal spectrum is scaled to the 90\% CL rejection sensitivity, as described in Sections~\ref{sec:stats} and \ref{sec:results}.  The right vertical axis specifies counts/keV in the 5.6~t 1000~d exposure, and is intended to give the reader a sense of the expected statistical fluctuations.}
\end{figure*}

Figure~\ref{fig:allSpec} describes the predicted spectra for the various signals considered.  The spectra of recoil energy (the true energy deposited) are shown by dashed curves in each plot.  As described in Section~\ref{sec:AnalFrame}, a NEST simulation then takes those true energies as inputs to generate the detector observables S1c and S2c.  These S1c and S2c signals can be combined to form a `reconstructed' energy as $E_{\rm{rec}} = W($S1c$/g_1 + $S2c$/g_2)$ where $W$ is the LXe work function, and $g_1$ and $g_2$ are the S1c and S2c signal gains as mentioned in Section II.  The spectra of this new $E_{\rm{rec}}$ quantity thereby include the effects of detector resolution and detector threshold.  Figure~\ref{fig:allSpec} shows signal model spectra in this $E_{\rm{rec}}$ quantity as solid lines.  Each $E_{\rm{rec}}$ signal spectrum is plotted using an amplitude corresponding to its specific rate sensitivity (90\% CL exclusion) as calculated in Section~\ref{sec:Sens}.  Finally, in each panel, the total ER background spectrum (in the $E_{\rm{rec}}$ quantity) is included for comparison. The right hand axis reports the predicted event counts per keV in the total expected LZ exposure (1000 live days operation).

\subsection{Electromagnetic properties of solar neutrinos: effective magnetic moment and millicharge}

Neutrinos are expected to exhibit non-zero electromagnetic couplings through loop contributions.  Such electromagnetic properties include an effective charge (or `millicharge') $q_\nu$ and an effective magnetic moment $\mu_\nu$.  In the Standard Model, minimally extended to include neutrino mass, the neutrino magnetic moment scales proportionally to neutrino mass as $\mu_\nu\approx3.2\times10^{-19} (m_\nu/1~\textrm{eV})~\mu_\textrm{B}$, where $m_{\nu}$ is a neutrino mass, $\mu_\textrm{B} = \frac{eh}{4\pi m_e}$ is the Bohr magneton, and $\mu_\nu$ is the neutrino magnetic moment of that mass state~\cite{fujikawa1980}.  Electromagnetic interactions at these small scales are beyond current experimental reach, but searches for neutrino magnetic moment and millicharge serve as useful tests of physics beyond the Standard Model.  Many extensions predict significant enhancement to neutrino electromagnetic properties~\cite{Bell_2005, Bell_2006, Giunti}. In the case that the neutrino is a Dirac particle, it is difficult to enhance the electromagnetic properties without a proportional increase to the neutrino mass, which is already tightly constrained.  This relationship with mass is less necessary in the Majorana case, and so any observation of neutrino electromagnetic properties would be a strong indicator of the neutrino being a Majorana particle.

Interactions via a millicharge or magnetic moment add terms to the total neutrino-electron scattering cross section, as
\begin{equation}
\begin{split}
\left(\frac{d\sigma_{\nu,e}}{dT_e}\right) \; \simeq  \; & \; \left(\frac{d\sigma_{\nu,e}}{dT_e}\right)_\textrm{weak} \\ & + \frac{\pi \alpha^2}{m_e^2} \left(\frac{1}{T_e} - \frac{1}{E_\nu}\right)\left(\frac{\mu_\nu}{\mu_\textrm{B}}\right)^2 \\ &
 + \frac{2 \pi \alpha}{m_e} \left(\frac{1}{T_e^2}\right)q_\nu^2 , 
\label{sigma_neutrino}
\end{split}
\end{equation}
where $\alpha$ is the fine structure constant, $m_e$ is the electron mass, and $T_e$ can be taken as the energy of the recoiling electron.  In the energy regime relevant to most experimental sensitivities, $T_e \ll E_\nu$, meaning the magnetic moment and millicharge scattering rates fall as $T_e^{-1}$ and $T_e^{-2}$ respectively. 

Eq.~\ref{sigma_neutrino} is a simplified description, assuming that the electron is unbound.  Atomic binding effects can be thought of, to first order, as reducing the number of electrons available for very low-energy scatters, thereby reducing the effective total scattering cross section and scattering rate at low energies.  The full calculation of atomic effects requires the RRPA method mentioned in Section IV, specific to the target atomic structure.  This calculation has recently been performed for neutrino electromagnetic scattering on Xe atoms by Hsieh~\emph{et al.}~\cite{Hseih_2019}, and it is these calculated spectra which serve as the starting point for our work, shown as the dashed lines in the upper middle panel of Figure~\ref{fig:allSpec}.  One can see both the overall $T_e^{-1}$ and $T_e^{-2}$ scaling, and the slight steps at the energies of the Xe electron shells.

Existing experimental sensitivities to neutrino electromagnetic properties vary by neutrino source and flavor mixture.  Reactor neutrinos fluxes, composed of pure $\bar\nu_e$, have been used by the GEMMA and TEXONO experiments to set upper limits of $\mu_{\bar\nu_e}<2.9\times 10^{−11}\mu_B$ and $q_{\bar\nu_e}<1.5\times10^{-12} e_0$, where $e_0$ is the positron charge~\cite{gemma,texono}.  Solar neutrinos ($\nu_{\textrm{solar}}$) arrive in terrestrial detectors as an incoherent mixture of the three flavor states, and electromagnetic interaction sensitivities can be quoted for the effective properties of that flavor mixture.  Panda-X-II has recently set an upper limit on the magnetic moment of $\mu_{\nu_\textrm{solar}}<4.9\times10^{−11} \mu_\textrm{B}$~\cite{PandaX_lowER}, while Borexino earlier set a stronger upper limit of $\mu_{\nu_\textrm{solar}}<2.8\times10^{−11} \mu_\textrm{B}$~\cite{borexino}.  XMASS-I has set an upper limit on the millicharge of $q_{\nu_\textrm{solar}}<5.4\times10^{-12}e_0$~\cite{xmass_2020}.  XENON1T has recently observed evidence of an excess that can be interpreted as a positive observation of $\mu_{\nu_\textrm{solar}}\in(1.4, 2.9)\times10^{-11}\mu_\textrm{B}$~\cite{xenon1t_2020}. \footnote{The XENON1T result emphasizes an important caveat, that if an unconstrained $^3$H background component is included, the statistical significance of the $\mu_{\nu_\textrm{solar}}$ evidence decreases from $3.2\sigma$ to $0.9\sigma$.}  This same XENON1T data could similarly be interpreted as a positive observation of $q_{\nu_\textrm{solar}}\in(1.7,2.3)\times10^{-12}e_0$~\cite{khan_2020}. 

As in other solar neutrino experiments, LZ will expose a massive low-background target to the solar neutrino flux.  As can be seen in Figure~\ref{fig:allSpec}, while the Standard Model solar neutrino recoils are not expected to dominate the rate, the sharply falling spectral shape that arises in the case of an electromagnetic enhancement, coupled to the $\sim$1~keV threshold, allows xenon detectors to compete with the much larger masses of dedicated neutrino observatories such as BOREXINO. This trade off between exposure and threshold is described in Ref.~\cite{Harnik_2012}.

\subsection{Solar axions}
The strong CP problem relates to the unnaturally small value of the CP violating term in QCD. 
The Standard Model can again be extended to provide a solution, for example, the Peccei--Quinn mechanism~\cite{PecceiQuinn_1977} introduces an additional U(1) chiral symmetry that is spontaneously broken. The CP violating term is then replaced by a dynamical field that automatically goes to zero by minimizing the potential. Axions are the hypothetical particle resulting from the field: Nambu-Goldstone bosons from the spontaneous breaking of this symmetry at some scale $f_a$. Couplings of axions to leptons, hadrons and photons are inversely proportional to $f_a$ and the smallest values ($f_a \leq 10^{-9}$~GeV) have been ruled out by experimental searches~\cite{Kim_2010}. Axion mass is also inversely proportional to $f_a$, therefore axions are expected to be very light and very weakly interacting. Production by a non-thermal realignment mechanism~\cite{Preskill_1983, Abbott_1983, Dine_1983} can generate sufficient quantities of cosmologically stable collisionless axions to make them a viable dark matter candidate. Further production in stellar environments is also possible, leading to emission from stars.

Axion-electron interactions would occur via the axio-electric effect, which is analogous to the photo-electric effect. This could produce ERs in LZ from solar axions emitted by the Sun. The solar axion flux, resulting from production of axions by reactions in the solar plasma, depends on the coupling of axions to electrons (coupling constant $g_\mathrm{Ae}$), photons ($g_\mathrm{Ag}$) and nucleons ($g_\mathrm{An}$). We consider three solar axion production mechanisms associated with these couplings:
\begin{enumerate}
\item Axion-electron coupling: Atomic, Bremsstrahlung and Compton (ABC)~\cite{Redondo_2013}
\item Axion-nucleon coupling: $^{57}$Fe de-excitation \cite{Moriyama_57FeAxion}
\item Axion-photon coupling: Primakoff effect \cite{Axion_Book}
\end{enumerate}

The relative importance of these production mechanisms depends on the axion model. In hadronic models, such as KSVZ \cite{KSVZ_Kim, KSVZ_Shifman}, the axion has no tree-level coupling to hadrons or leptons, so the Primakoff production dominates. However, for general axion models, such as DFSZ \cite{DFSZ_Dine, DFSZ_Zhitnitsky}, the ABC processes, due to the electron coupling, will dominate production. Here no model is assumed and the three components are treated independently of one another.

The ABC reactions, driven by the axion-electron coupling, comprise: atomic axio-recombination and axio-deexcitation (A), electron ion and electron electron bremsstrahlung (B) and Compton scattering (C). Despite A having the largest cross section it only contributes for metal ions, which are less abundant in the Sun than hydrogen, helium and electrons; therefore B dominates. The flux is found by integrating the sum of emission rates from ABC, multiplied by the phase space density, over the volume of the Sun - tabulated values for this flux are given in Ref. \cite{Redondo_2013}.

The axion-nucleon coupling can lead to axion emission from M1 transitions as a result of de-excitation from thermally excited nuclides within the Sun. This requires an accessible first excited state with an M1 transition to the ground state, in addition to having a high natural abundance in the Sun. The $^{57}$Fe isotope fulfils these conditions: it has a 14.4~keV first excited state that may be populated given the solar temperature $\sim$1.3~keV, and has a 2.6\% solar abundance by mass fraction. The state is known to de-excitate via M1 transitions to the ground state, emitting $\gamma$-rays and internal conversion electrons. De-excitation through the emission of
a 14.4~keV axion would produce a flux \cite{CAST_57Fe}
\begin{equation}
\Phi_a =  \Big(\frac{k_a}{k_{\gamma}}\Big)^3 \times  (g_{\rm{An}}^{\rm{eff}})^2 \quad [4.56 \times 10^{23} \; \rm{cm}^{-2} \rm{s}^{-1}],
\end{equation}
where $k_a$ and $k_\gamma$ are the momenta of the axion and photon, and $g_{\rm{An}}^{\rm{eff}}$ is the effective axion nucleon coupling.

The axion-photon coupling drives Primakoff production --- particles with a two photon vertex can be produced from thermal photons in an external field. The strong magnetic fields and thermal photons in the Sun enable this production mechanism, with flux \cite{Axion_Book}
\begin{equation}
\frac{d\Phi}{dE} = g_{10}^2 \; \textit{E}\; {2.481}e^{-\textit{E}/1.205} \; [ 6 \times 10^{10} \; \rm{cm}^{-2}\rm{s}^{-1} \rm{keV}^{-1} ],
\end{equation}
where $g_{10} = g_{A\gamma} / 10^{-10} $ GeV$^{-1}$.

All of these flux components must be multiplied by the axio-electric cross section and the xenon atomic number density in order to find the expected rate in LXe. The solid line in the lower middle panel of Figure~\ref{fig:allSpec} shows the expected energy spectrum in $E_{\rm{rec}}$ for the ABC component. Event rates are again dominated by the near-threshold behavior and the experimental resolution smooths out atomic shell effects.

\subsection{Axion-like particles}\label{sec_ALP}
Axion-like particles (ALPs) are a general type of massless Nambu-Goldstone boson or massive pseudo Nambu-Goldstone boson that appears in many BSM models as a result of spontaneous breaking of additional (and/or approximate) global symmetries. Although they share some qualitative properties, couplings of generic ALPs to Standard Model particles are far less constrained than for axions, i.e. ALP masses and their coupling to photons are not related and ALPs are not linked to the Peccei–Quinn mechanism in QCD. Thus, while axions acquire mass from mixing with neutral pseudoscalar mesons ($m_{\rm{A}} f_{\rm{A}} \sim m_\pi f_\pi$), ALPs do not, i.e. the  quantity $m_{\rm{ALP}}f_{\rm{ALP}}$ is less restricted than for QCD axions. This makes the parameter space for ALPs much wider in the experimental context. 

Absorption of an ALP by a bound electron is analogous to the ordinary photo-electric effect, with the photon energy $\omega$ replaced by the ALP rest mass $m_{\rm{ALP}}$.  The cross-section of an ALP absorption (i.e. the axio-electric effect) can be expressed in terms of the photo-electric cross-section $\sigma_{\rm{PE}}$($\omega=m_{\rm{ALP}}$) as~\cite{PhysRevD.78.115012}
\begin{equation}
\frac{\sigma_{\rm{ALP}} v}{\sigma_{\rm{PE}}(\omega=m_{\rm{ALP}}) c} \approx \frac{3m^2_{\rm{ALP}}}{4\pi\alpha f^2_\alpha},
\label{CS_ALP}
\end{equation}where $f_\alpha \equiv 2m_{\rm{e}}/g_{\rm{Ae}}$ and $g_{\rm{Ae}}$ are the dimensional and dimensionless coupling constants respectively. $m_{\rm{ALP}}$ is the mass and $v \sim 10^{-3}c$ is the velocity of the ALP CDM. The expected event rate (kg$^{-1}$ day$^{-1}$) in the detector is~\cite{PhysRevD.78.115012}
\begin{equation}
R_{\rm{ALP}} \simeq \frac{1.2 \times 10^{19}}{A} g^2_{\rm{Ae}} {\sigma}_{\rm{PE}}{\,{m}_{\rm{ALP}}},
\label{ALP_R}
\end{equation}where $\sigma_{\rm{PE}}$ is expressed in barns and $m_{\rm{ALP}}$ in keV/c$^2$. The rate $R_{\rm{ALP}}$ is independent of the velocity $v$, similar to that for the hidden photo-electric effect discussed in Section~\ref{HPsection}.

In the axio-electric effect, the entire ALP rest mass is converted into energy and absorbed by the atomic electron, i.e. the energy deposition is essentially equal to the incoming mass in keV/c$^2$. The expected energy deposition spectrum in the detector is a mono-energetic peak at the value of the incident mass (then smeared by the experimental resolution, as described in Section~\ref{sec:AnalFrame}). The mass range scanned over thus defines an equivalent range of energy deposition in the detector. In this analysis, the lowest ALP mass considered is 2.0~keV, limited by the ER energy threshold of $\sim$1~keV. The largest ALP mass considered here is 70~keV, though in principle there is no reason this range could not be extended to higher masses in a later analysis.  Two example ALP signal models (for m$_{\rm{ALP}}$=20~keV and m$_{\rm{ALP}}$=60~keV) are shown in the lower right plot in Figure~\ref{fig:allSpec}.

\subsection{Hidden/dark photons}\label{HPsection}
The hidden (or dark) photon (HP) is a hypothetical U(1)$'$ gauge boson that resides in a hidden sector, i.e. a sector that does not interact with Standard Model particles through known Standard Model forces. Extra hidden U(1)$'$ symmetries often appear in supersymmetric extensions of the Standard Model and in string theories. Hidden photons can obtain a mass either via a hidden Higgs or a St\"uckelberg mechanism, and interact with the visible sector via loop-induced kinetic mixing~\cite{Abel:2008ai} with Standard Model hypercharge U(1)$_Y$ gauge bosons. If non-thermally produced via the misalignment mechanism in the early universe they can reproduce the present day dark matter relic abundance~\cite{Arias_2012}. 
As for the absorption of ALPs, the absorption of a HP by a bound electron is analogous to the ordinary photo-electric effect, with the photon energy $\omega$ replaced by the HP rest mass $m_{\rm{HP}}$. The two cross-sections, $\sigma_{\rm{HP}}$ and $\sigma_{\rm{PE}}$, are related by~\cite{PhysRevD.78.115012}
\begin{equation}
\frac{\sigma_{\rm{HP}} v}{\sigma_{\rm{PE}}(\omega=m_{\rm{HP}}) c} \approx \frac{\alpha^\prime}{\alpha},
\label{CS_HP}
\end{equation}where $v$ is the velocity of the HP CDM particle, $\alpha$ is the electromagnetic fine structure constant and $\alpha^\prime$ is its analogue for HP
\begin{equation}
\alpha =\frac{e^2}{4\pi}, \qquad \alpha^\prime =\frac{g_{\rm{h}}^2}{4\pi}.
\label{FSC}
\end{equation}
$e$ and $g_{\rm{h}}$ are the visible and gauge couplings respectively. The kinetic mixing parameter $\kappa$ is related to the fine structure constants $\alpha$ and $\alpha^\prime$ as~\cite{Brummer:2009cs}
\begin{equation}
\kappa = \bigg(\frac{\alpha^\prime}{\alpha}\bigg)^{1/2}.
\label{HP_kappa_Chap6}
\end{equation}

For HPs to constitute cold dark matter they must have a density of 0.3~GeV/cm$^{3}$ and be  non-relativistic. This corresponds to $v\sim$10$^{-3}c$, and the expected interaction rate (kg$^{-1}$ day$^{-1}$) in the detector in this approximation follows Ref.~\cite{PhysRevD.78.115012}
\begin{equation}
R_{\rm{HP}} \simeq \frac{4 \times 10^{23}}{A} \frac{\alpha^\prime}{\alpha} \frac{\sigma_{\rm{PE}}}{m_{\rm{HP}}},
\label{HP_R}
\end{equation}
where $\sigma_{\rm{PE}}$ is expressed in barns and $m_{\rm{HP}}$ in keV/c$^2$. $A=131.3$ is the atomic mass of Xe. The event rates are independent of the HP velocity distribution in the galactic halo, suggesting the absence of any annual modulation by the Earth's motion, at least to an experimentally relevant level~\cite{PhysRevD.78.115012}.

As for the ALP case (Section~\ref{sec_ALP}), the expected energy deposition spectrum in the detector should be a mono-energetic peak centred at the value of the incident HP mass and smeared by the experimental resolution. Example energy deposition spectra for 20~keV/c$^{2}$and 60~keV/c$^{2}$ HP or ALP are shown in the lower right plot of Figure~\ref{fig:allSpec}, again at the amplitudes of projected 90\% CL rejection sensitivity according to the profile likelihood ratio tests of Section VI and VII.

\subsection{Mirror Dark Matter}
Whilst hidden photons (discussed above) are a general feature of any hidden sector model with a U(1)' group, there are also models where the hidden sector has a specific structure. An example is the mirror dark matter (MDM) model, where the hidden sector is isomorphic to the SM~\cite{IntJModPhysA.29.0217}. 
This symmetry means the hidden sector contains a mirror partner of each Standard Model particle, with the same masses, lifetimes and self interactions. The Lagrangian for the system contains the Standard Model Lagrangian, a mirror analog and a term describing mixing/portal interactions between the two,

\begin{equation}
\Lagr_{mix} = \frac{\varepsilon}{2} F^{\mu \nu} F'_{\mu \nu} + \lambda \phi^{\dagger} \phi \phi^{' \dagger} \phi^{'}.
\end{equation}
The first term describes kinetic mixing of $U(1)_Y$ and mirror $U(1)_Y'$ and the second describes Higgs ($\phi$) -- mirror Higgs ($\phi'$) mixing. Of interest here is the kinetic mixing interaction with strength $\varepsilon$~\cite{PhysLettB.272.67}, which induces a tiny $\varepsilon e$ electromagnetic charge in the mirror particles charged under U(1)'.

MDM would exist as a multi-component plasma halo with the main components being mirror electrons, $\rm e'$, and mirror helium nuclei, $\rm He'$. Kinetic mixing between mirror and Standard Model particles would allow mirror electrons in the halo to scatter off Xe atomic electrons in the LZ detector, inducing ERs.

The rate of ERs depends on the kinetic mixing strength $\varepsilon$ and the local mirror electron temperature $T$. There is also a non-negligible terrestrial effect from the capture and subsequent shielding of incoming mirror electrons \cite{PhysLettB.789.592}.  This alters the incoming flux and velocity distributions, as calculated in Ref. \cite{PhysRevD.101.012003}.

The expected rate is given by \cite{PhysLettB.789.592},  
\begin{equation} \label{eq:diffrate}
\begin{split}
\frac{dR}{dE_R} = g_T N_T n_{e'}^0 \frac{\lambda}{v_c^0 E_R^2} [1 + & A_v\rm cos\, \omega(t - t_0) \\ & +  A_{\theta} (\theta - \bar{\theta})],
\end{split}
\end{equation}
where $N_T$ is the number of target electrons and $n_{e'}^0$ is the number density of mirror electrons arriving at the detector. The effective number of free electrons, $g_T$, is the number of electrons per target atom with atomic binding energy ($E_b$) less than the recoil energy ($E_R$), modelled as a step function for the atomic shells in xenon. The energy dependent $v_c^0$ term describes the modified velocity distribution due to terrestrial effects.

Significant modulation, both annual and sidereal, is expected for plasma dark matter models, including MDM. The $A_v$ term describes annual modulation due to a Galilean boost from the variation in speed of the Earth with respect to the MDM halo, with $A_v$ the amplitude, $\omega$ the angular frequency, $t$ the day for which the rate is calculated, and $t_0$ the day with maximum speed with respect to the halo (day 152, 2nd June). The $A_{\theta}$ term describes annual and sidereal modulation due to variation of the angle between the Earth's spin axis and the incoming dark matter wind: $A_{\theta}$ is the amplitude and $\theta(t)$ is the angle between the halo wind and the zenith, with $\bar{\theta}$ being the average over a year.

These date dependent modulation terms are calculated for a given date, with maximum daily variations in rate of approximately 60\% on the 7th June and 8th December. These variations average out over the course of a year and for 1000 live days we find a maximum variation of 7\%. However there may be greater variation depending on any gaps in data-taking. Since the exact dates are not known yet, a factor of 1.0 is used for these two terms in the rate calculations.

The resulting theoretical energy spectra and corresponding reconstructed energy spectra 
are shown in the upper left panel of Figure~\ref{fig:allSpec} for two mirror electron temperatures with the kinetic mixing parameter at the sensitivity level determined in Section~\ref{sec:Sens}. 

\subsection{Leptophillic EFT}
Lastly, we consider a model in which the dark matter particle couples only to leptons, via an axial-vector interaction. In this case, the signal of interest to LZ is generated by WIMPs scattering off electrons, such that the recoil energy acquired by the electron is sufficient to liberate it from its host atom. While the most general tree-level effective interaction between the dark matter particle and leptons allows for loop-induced interactions with quarks that then dominate the experimental signature, considering only the axial-vector interactions results in the loop contributions vanishing~\cite{leptophilicDM2014}.

At typical WIMP velocities, the energy of recoiling electrons that were initially at rest is of the order of a few eV and therefore far below the detection threshold. However, an electron bound to an atom can have a non-negligible momentum, resulting in detectable energy depositions. The energy spectrum expected at the detector can be determined as described in Eq. 30 in Ref. \cite{leptophilicDM2009} by integrating over the momentum wave function of each electron shell and summing over shells. This describes the contribution from each shell at any given recoil energy and imposes the requirement that the recoil energy exceeds the binding energy of the given shell. The lower left plot in Figure \ref{fig:allSpec} shows the expected event rate in LZ,  with the dominant contribution to the signal arising from scattering off electrons in the 3s shell.

\section{Statistical analysis}\label{sec:stats}
\noindent
We employ frequentist hypothesis tests based on the PLR method, and follow the prescription of Ref.~\cite{Rolke_paper} to determine LZ's sensitivity to exclude or observe parameters of the models described above.

The comparison between signal and background is conducted using two-dimensional probability distribution functions (PDFs) in the variables S1c and log(S2c). The choice to use 2D PDFs in these observables offers some small advantages in sensitivity compared to the simpler analysis using $E_{\rm{rec}}$. The PDFs are produced in NEST using the energy spectra of the signal and backgrounds as in Figure~\ref{fig:allSpec}.
An extended unbinned likelihood function is then generated with Gaussian-constrained nuisance parameters for the dominant background rates.  

The projected exclusion sensitivity reported for a given signal model is the median 90\% CL upper limit on the model's strength parameter-of-interest that would be obtained in repeated background-only experiments. We use a two-sided PLR test statistic~\cite{cowan, stats_whiteppaer}, the distributions of which are obtained through toy Monte Carlo (MC) experiments generated from the likelihood model (as opposed to relying on asymptotic approximations of the distribution). A total of 5000 toy experiments are used to build the test statistic distributions under both the null and alternate hypothesis for each parameter-of-interest value scanned.

Exclusion sensitivities are calculated for all signal models described in Section~\ref{sec:signals}. Additionally, for those signal models deemed most relevant to the excess reported by XENON1T~\cite{xenon1t_2020} we also calculate the median parameter-of-interest values at which the background-only model is rejected at 3$\sigma$ ($p$-value threshold $1.3\times10^{-3}$), indicating evidence of a signal. These selected models are: the solar neutrino magnetic moment, solar neutrino millicharge, solar axions, and the monoenergetic signals from ALPs and HPs.

\begin{table}[]
\caption{\label{tab:backgrounds} We list the expected mean counts for each background component in a 5.6~t 1000~d exposure for the specified ranges of energy and S1c.  Counts are quoted after the application of single scatter, veto, and fiducial volume selections. The last column specifies the uncertainty in rate we assume when projecting expected sensitivity in Section~\ref{sec:stats}.  Solar neutrino electron scattering is included without uncertainty.  Nuclear recoil backgrounds are included in the analysis (see ~\cite{LZ_wimp}) but are omitted from this table due to their negligible rates within the ER signal region.}
\begin{tabularx}{\columnwidth}{lXXXXc}
\hline \hline \rule{0pt}{2ex}
 & \multicolumn{4}{l}{\textbf{Expected counts in 5.6~t 1000~d}} &          \\  
Component & \multicolumn{2}{l}{Energy range [keV]} & \multicolumn{2}{l}{S1c range [phd]}  &  Unc. \\
                   & 1.5-10      & 1.5-70     & 0-100     & 0-570     &  (\%)               \\ \hline \rule{0pt}{3ex}
$^{222}$Rn         & 1216           & 9873          & 2504         & 11998        & \multirow{3}{*}{24} \\
\kern 0.2pc $^{220}$Rn         & 171            & 1394           & 353          & 1694         &                       \\
\kern 0.5pc $^{85}$Kr          & 45             & 378            & 93           & 462          &                       \\ \rule{0pt}{4ex}
$^{136}$Xe         & 166            & 8796          & 603          & 13186        & 50                   \\ \rule{0pt}{5ex}
$^{124}$Xe         & 38            & 3287          & 56          & 3299        & 30                   \\  \rule{0pt}{5ex}
Solar $\nu$        & 336            & 2418           & 670          & 2845         &  ---                  \\  \rule{0pt}{5ex}
Det.           &                &                &              &              &                       \\
+Surf.           & 93             & 754            & 191          & 916          & 20                   \\
+Env. $\gamma$     &                &                &              &              &                     \\ \hline \rule{0pt}{2.5ex}
 TOTAL          & 2065            & 26900           & 4470          & 34400         &                       \\   
\hline \hline 
\end{tabularx}
\end{table}

The expected rates and uncertainties for the background components used in the likelihood are listed in Table~\ref{tab:backgrounds}. Signals from decays of $^{85}$Kr, $^{222}$Rn, and $^{220}$Rn (and importantly, their progeny $^{214}$Pb and $^{212}$Pb) are grouped together as a single background component with a single, combined rate uncertainty due to their similar spectral shapes in the region of interest.  Two background components are included without rate uncertainties, and their normalizations are fixed in the analysis. The first is solar neutrino-electron recoils, the rate of which is dominated by neutrinos from the {\it pp} chain. The corresponding flux uncertainty is estimated to be 0.6\%~\cite{Vinyoles2017}.  The second is the general category of NR backgrounds, due to the negligible number of which are expected in the ER signal region in \{S1c, log(S2c)\} space, as mentioned previously.  In the solar axion analysis, the Primakoff component of the signal model was used as an additional, unconstrained nuisance parameter that floats in the fit.   

As mentioned in Section~\ref{sec:backgrounds}, we include two `unexpected but possible' backgrounds in the specific case of projecting discovery sensitivity:  $^{37}$Ar and $^3$H.  In a real experimental dataset, evidence for new physics would not be claimed if the observed excess were similarly consistent with some unexpected but possible background contamination.  Because they are unexpected, these two backgrounds are not included in the generation of toy MC data, but because they are possible, they are included in the likelihood fits to those data.  In a real experiment, some externally-derived constraint on $^{37}$Ar and $^{3}$H concentrations or rates may be applied, but in this work we conservatively apply no such constraint.  The result of this method is that no evidence for a discovery is achieved for signals with spectral shape identical to that of either $^{37}$Ar or $^{3}$H, and sensitivity is reduced for any signal with sufficiently high overlap with either of these two backgrounds.

\section{Results}\label{sec:results}

\subsection{Sensitivity projections}
\label{sec:Sens}

\noindent
We summarize the results of the sensitivity projections, starting with the case of neutrino electromagnetic properties.  We find that the full LZ exposure will be capable of rejecting solar neutrino magnetic moments $\mu_{\nu_\textrm{solar}}$  greater than ${6.2 \times 10^{-12} \mu_{\rm{B}}}$ and millicharges $q_{\nu_\textrm{solar}}$ greater than ${1.4 \times 10^{-13} e_0}$, both at 90\% CL.  Sensitivity to 3$\sigma$ evidence is found to be ${9.5 \times 10^{-12} \mu_{\rm{B}}}$ and ${1.9 \times 10^{-13} e_0}$.  To re-emphasize, these sensitivities to evidence conservatively include unconstrained rates of the unexpected $^{37}$Ar and $^{3}$H backgrounds, weakening discovery sensitivity. 

\begin{figure*}[pht!]

\begin{subfigure}{\columnwidth}
\includegraphics[width=\columnwidth]{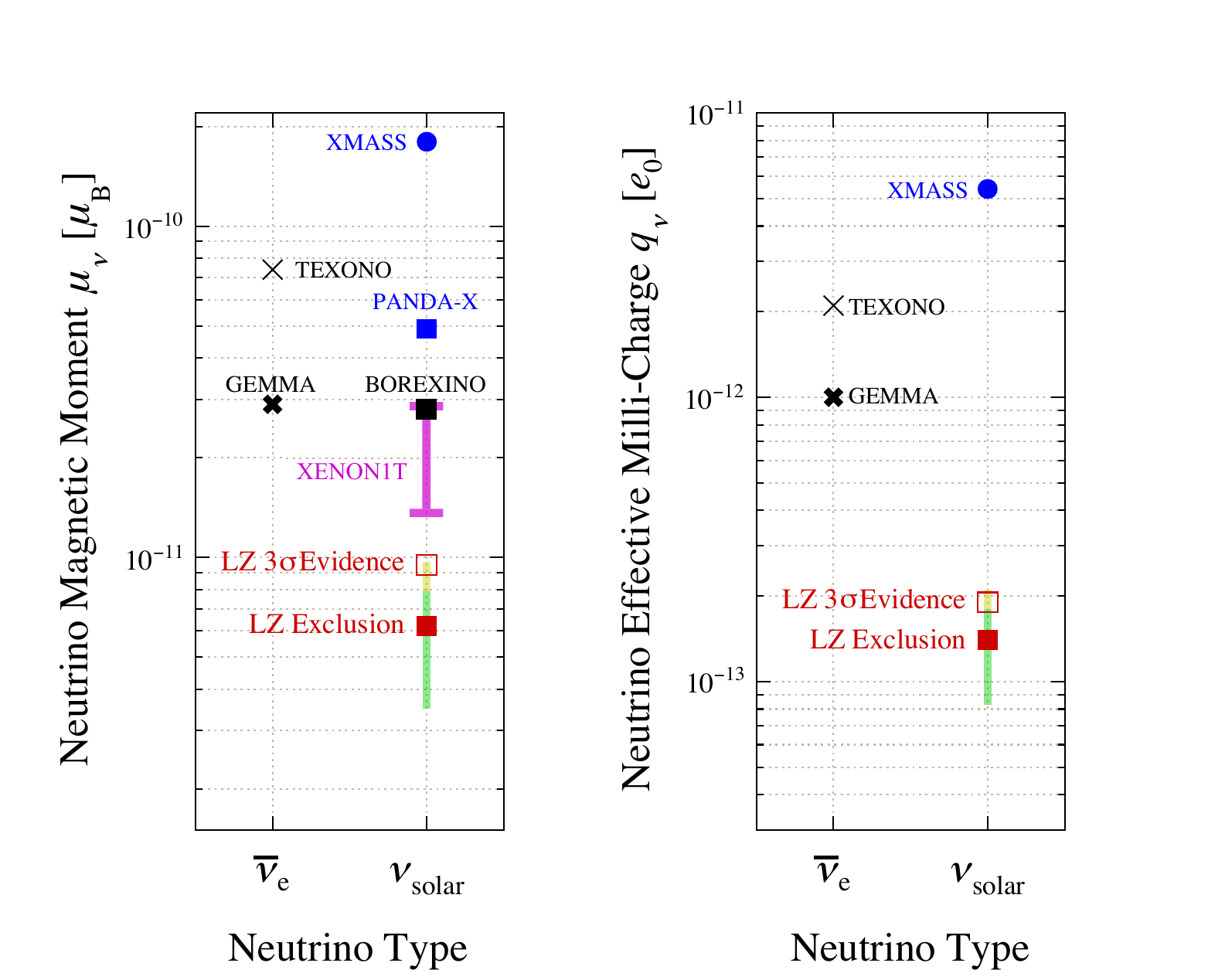}
\caption []{\label{fig:nuSens}}
\end{subfigure}
\begin{subfigure}{\columnwidth}
\includegraphics[width=\columnwidth]{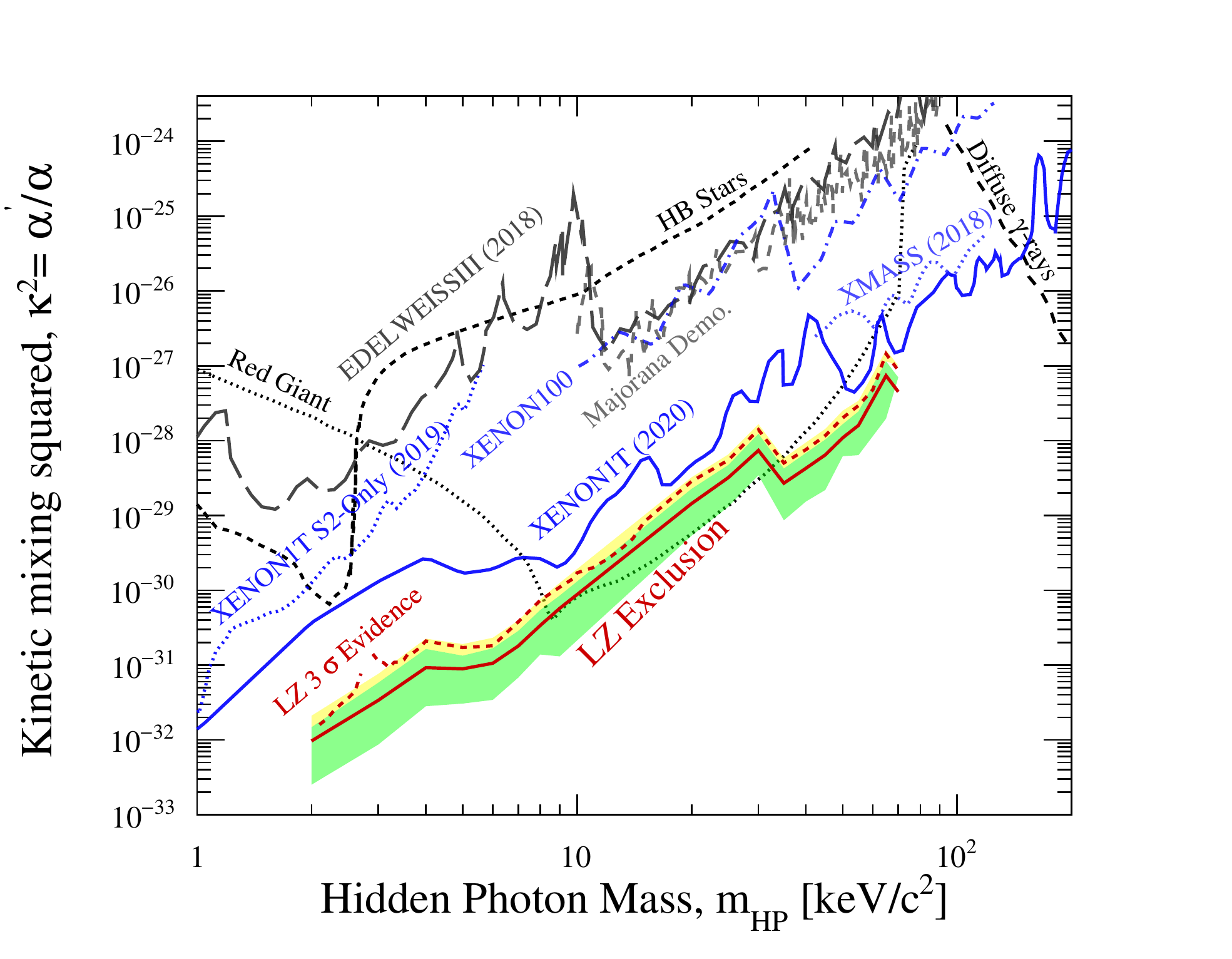}
\caption []{\label{fig:HPSens}}
\end{subfigure}
\begin{subfigure}{\columnwidth}
\includegraphics[width=\columnwidth]{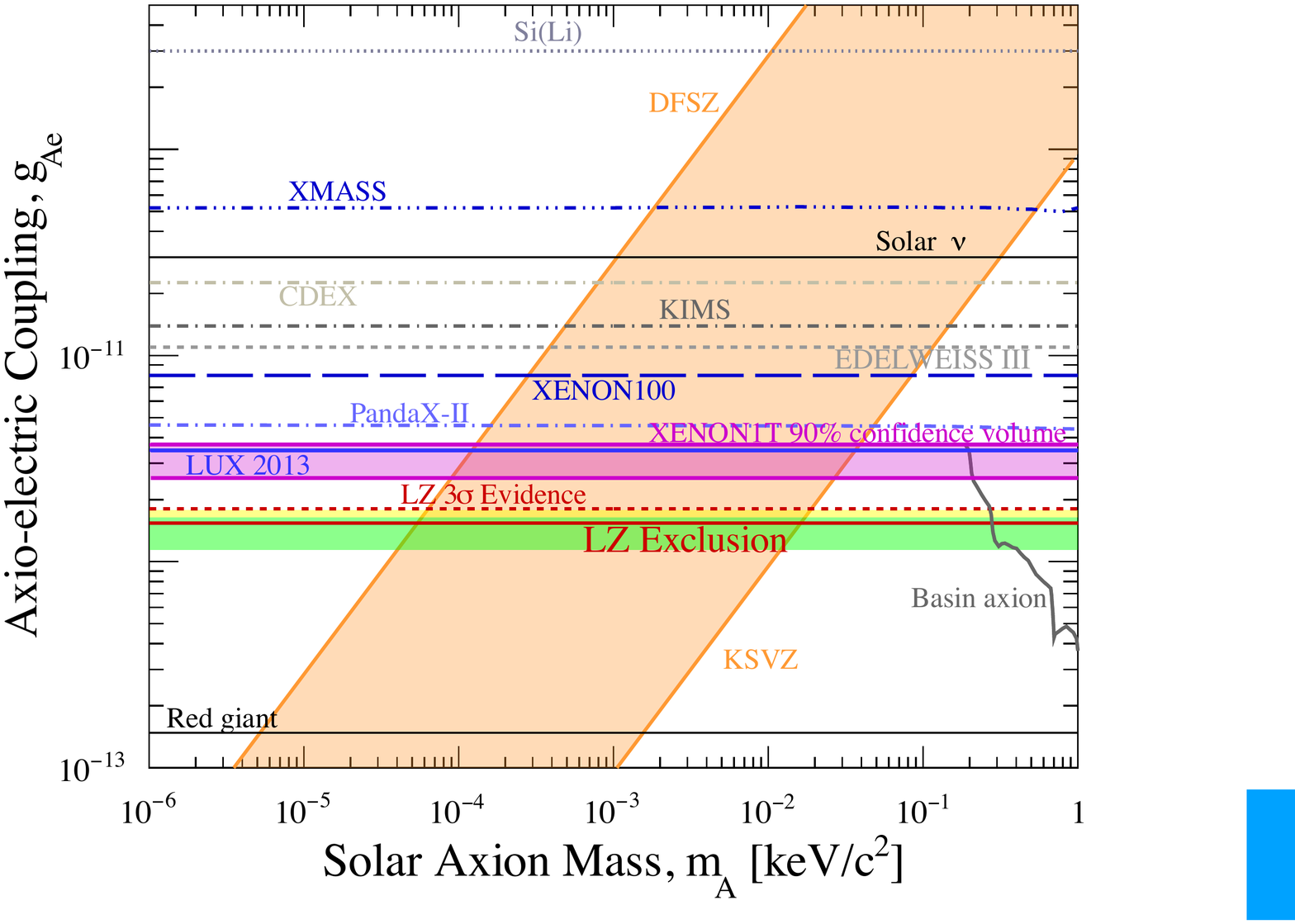}
\caption []{\label{fig:axionSens}}
\end{subfigure}
\begin{subfigure}{\columnwidth}
\includegraphics[width=\columnwidth]{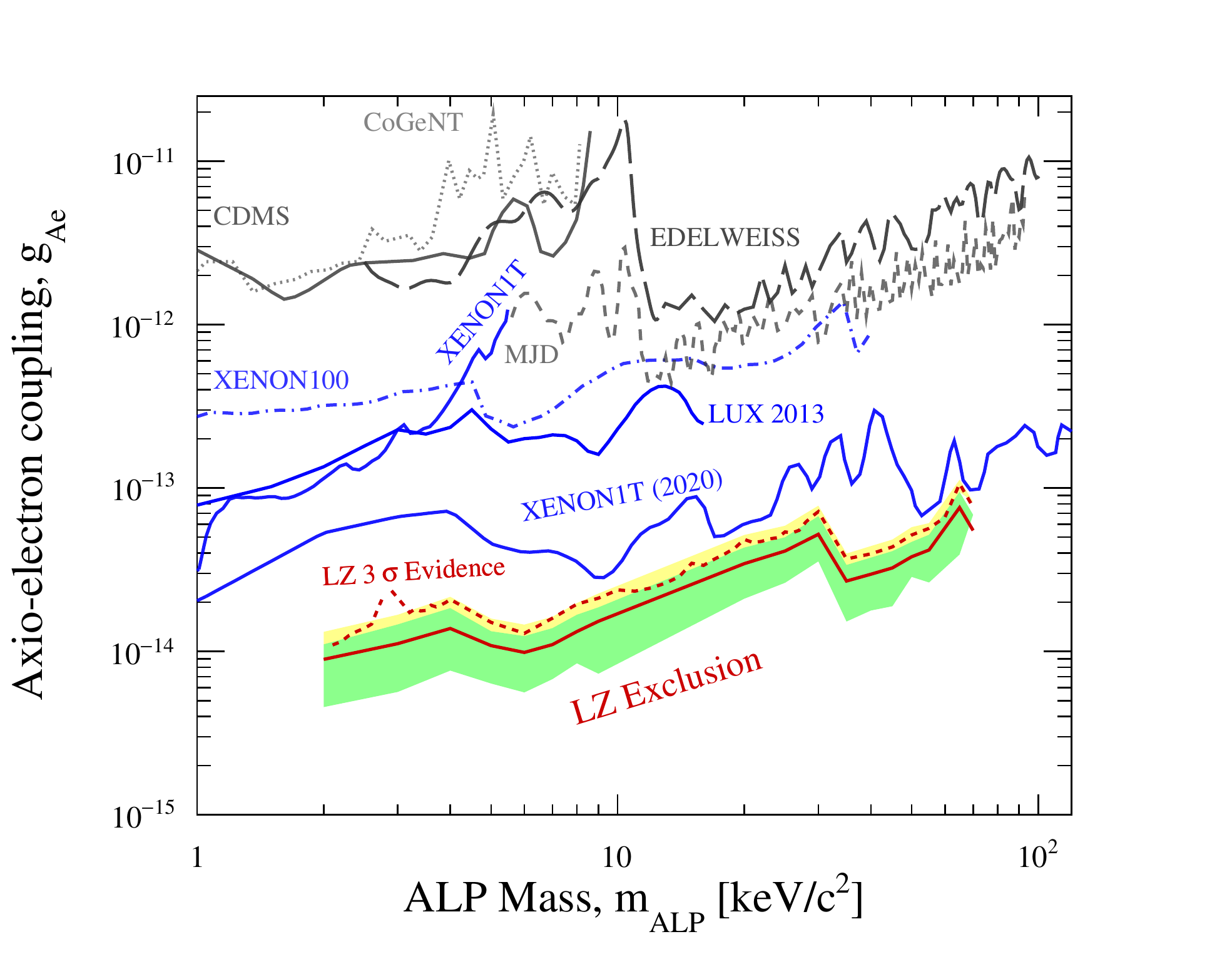}
\caption []{\label{fig:APSens}}
\end{subfigure}
\begin{subfigure}{\columnwidth}
\includegraphics[width=0.92\columnwidth]{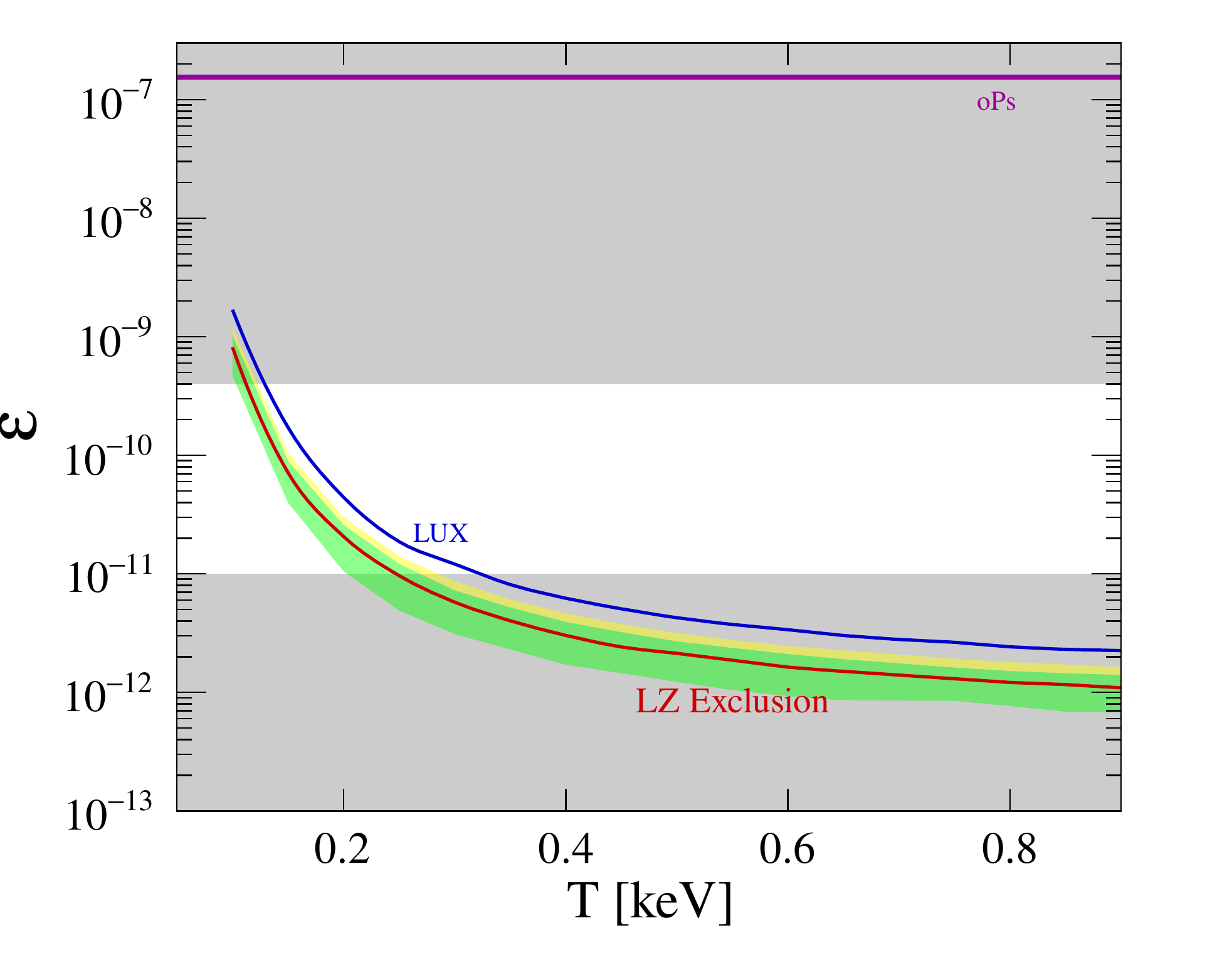}
\caption []{\label{fig:mdmSens}}
\end{subfigure}
\begin{subfigure}{\columnwidth}
\includegraphics[width=\columnwidth]{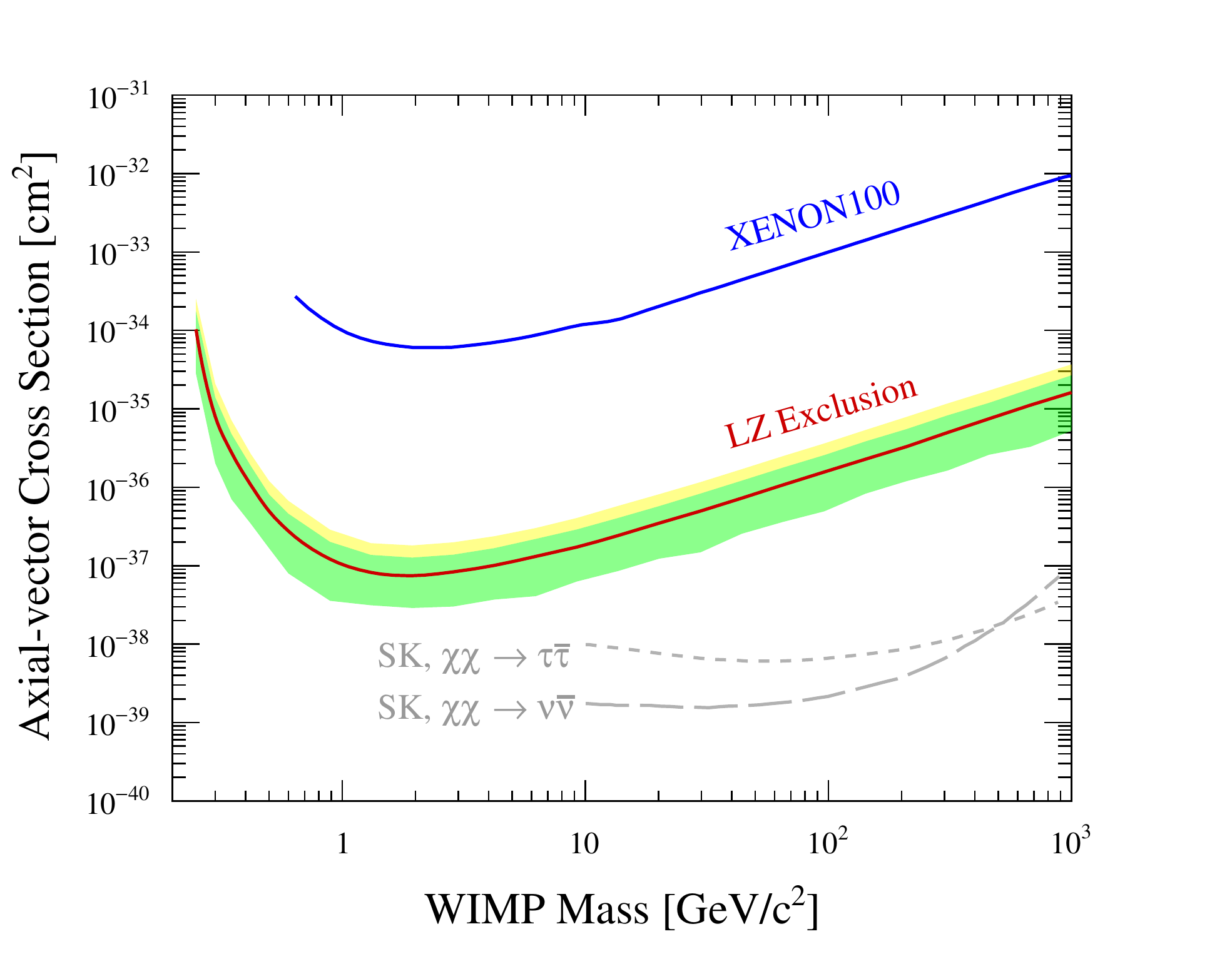}
\caption []{\label{fig:lepto}}
\end{subfigure}
\caption{Projected $90\%$ C.L. exclusion sensitivity to (\subref{fig:nuSens}) electromagnetic neutrino couplings $\mu_\nu$ and $q_\nu$, (\subref{fig:HPSens}) kinetic mixing squared, $\kappa^2$ for hidden photons, (\subref{fig:axionSens}) axio-electric coupling for solar axions , (\subref{fig:APSens}) axio-electric coupling for galactic ALPs,    (\subref{fig:mdmSens}) mirror dark matter kinetic mixing, and (\subref{fig:lepto}) axial-vector cross-section for leptophilic dark matter.  $\pm 1\sigma$ (green) and $+2\sigma$ (yellow) bands are also shown. For selected models, sensitivity to 3$\sigma$ evidence is also shown.  Results from other experiments and astrophysical constraints are referenced in the text.}
\label{fig:SensMono}
\end{figure*}

These sensitivities are shown in Figure~\ref{fig:nuSens}, together with existing experimental constraints on solar neutrino properties.  Constraints using pure $\bar{\nu}_e$ neutrinos (from reactor-based experiments) are also shown.  XENON1T has seen evidence for an excess in rate at low energies consistent with several signal hypotheses including a neutrino magnetic moment~\cite{xenon1t_2020}.  The 90\%~CL allowed region under this magnetic moment interpretation is indicated.  We see from the LZ discovery and exclusion sensitivities that the XENON1T excess, under the magnetic moment hypothesis, will be robustly tested by LZ.  We note that various astrophysical constraints are stronger than these direct experimental constraints, see for example a summary in \cite{Giunti}.  Given the sharply falling spectral shape, particularly in the millicharge case, an S2-only treatment of the LZ data may have a similar or stronger sensitivity to these signals, but the backgrounds to such searches are difficult to estimate \emph{a priori} and the topic is beyond the scope of this work.

For solar axions we find projected 90\%~CL exclusion sensitivity to axio-electric coupling $g_{\rm{Ae}}$ values as low as 1.58$\times10^{-12}$, and 3$\sigma$ evidence for discovery for $g_{\rm{Ae}}$ equal to 1.84$\times$10$^{-11}$.  These sensitivity projections are shown in Figure~\ref{fig:axionSens}. This shows an improvement over previous experimental results, including a factor $\sim$2.5 in exclusion sensitivity over the LUX result~\cite{LUX_axion}. The most stringent limit at present is the  indirect constraint derived from the observed brightness of the tip of the Red Giant branch~\cite{RGaxion_2020}, which in turn places limits on the allowed additional cooling by axion emission. The indirect constraint from the solar neutrino flux~\cite{Gondolo_nuaxion} and experimental constraints from both solid state~\cite{Derbin_axion, EDELWEISS_ER, KIMS_axion, CDEX_axion} and LXe direct detection~\cite{XMASS_axion, XENON100_axion, PandaX_axion, PandaX_lowER} experiments are also presented. Also shown in Figure~\ref{fig:axionSens} is a constraint from a model that includes a ``stellar basin'' of gravitationally bound axions, giving a second component of the flux~\cite{Basin_axion}. This result is obtained by recasting previous experimental limits, but there is a large uncertainty from calculation of the ejection time of particles from the solar system.  The XENON1T excess~\cite{xenon1t_2020} can be interpreted as an allowed range of $g_{\rm{Ae}}$ for the case of other couplings (e.g. $g_{\rm{A\gamma}}$) being small, and that allowed region in $g_{\rm{Ae}}$ is shown in Figure~\ref{fig:axionSens}. LZ will robustly test a solar axion interpretation of the XENON1T excess.

A scan over ALP masses (2--70~keV/c$^{2}$) constrains the expected $g_{\rm{Ae}}$ to be no larger than $\sim$7.6$\times 10^{-14}$ (at $\sim$64.3~keV, the energy of the $^{124}$Xe KK peak). LZ is expected to give a limit approximately two orders of magnitude more stringent than existing results~\cite{2018153,Aprile:2019xxb,An:2020bxd}.
A scan over HP masses (2--70~keV/c$^{2}$) constrains $\kappa^2$ to be no larger than $\sim$7.4$\times 10^{-28}$ (at $\sim$64.3~keV). While at intermediate energies ($\sim$8-30~keV) the indirect (Red Giant) limit on $\kappa^2$ is still the most stringent, LZ is expected to give a better limit at lower ($\leqslant$8~keV) and higher energies, i.e. $\sim$(30--70)~keV. An improvement of approximately two orders of magnitude over the existing direct detection results~\cite{2018153,Aprile:2019xxb,An:2020bxd} is expected.  Because an unconstrained $^{37}$Ar component is included in projections of sensitivity to signal evidence, we expect no sensitivity to monoenergetic signals at the 2.8~keV energy of the $^{37}$Ar decay.  Note the discontinuity at this energy in both the ALP and HP evidence sensitivity curves (dashed red curves in Figures~\ref{fig:APSens} and \ref{fig:HPSens}).

The projected exclusion sensitivity to mirror dark matter kinetic mixing for various local mirror electron temperatures is shown in Figure~\ref{fig:mdmSens}. This shows that, in the event of no signal, we would be able to exclude mirror electron temperatures down to 0.25~keV for this model, giving an improvement over the first direct detection search carried out by LUX~\cite{PhysRevD.101.012003}.

The projected exclusion sensitivity to the axial-vector WIMP-electron scattering cross section is as low as $7.5\times10^{-38}~\mathrm{cm}^2$ for a WIMP mass of 2~GeV/c$^2$, as shown in Figure~\ref{fig:lepto}. This represents an improvement of approximately three orders of magnitude over the strongest direct detection limit to date, set by XENON100 in~\cite{XENON100leptophilic}. While this will allow LZ to probe new parameter space for WIMP masses below 10~GeV/c$^2$, indirect constraints on DM annihilation in the sun from Super-Kamiokande remain the most stringent for higher masses~\cite{leptophilicDM2009}. We note that a lower bound on the coupling from relic density constraints implies an upper limit on the axial-vector cross section of $\sim$10$^{-47}$~cm$^2$, well beyond the reach of current generation experiments \cite{leptophilicILC}.

\subsection{Effect of intrinsic beta backgrounds}\label{sec:variedbetas}
\begin{figure*}[pht!]
\begin{subfigure}{\columnwidth}
\includegraphics[width=\columnwidth]{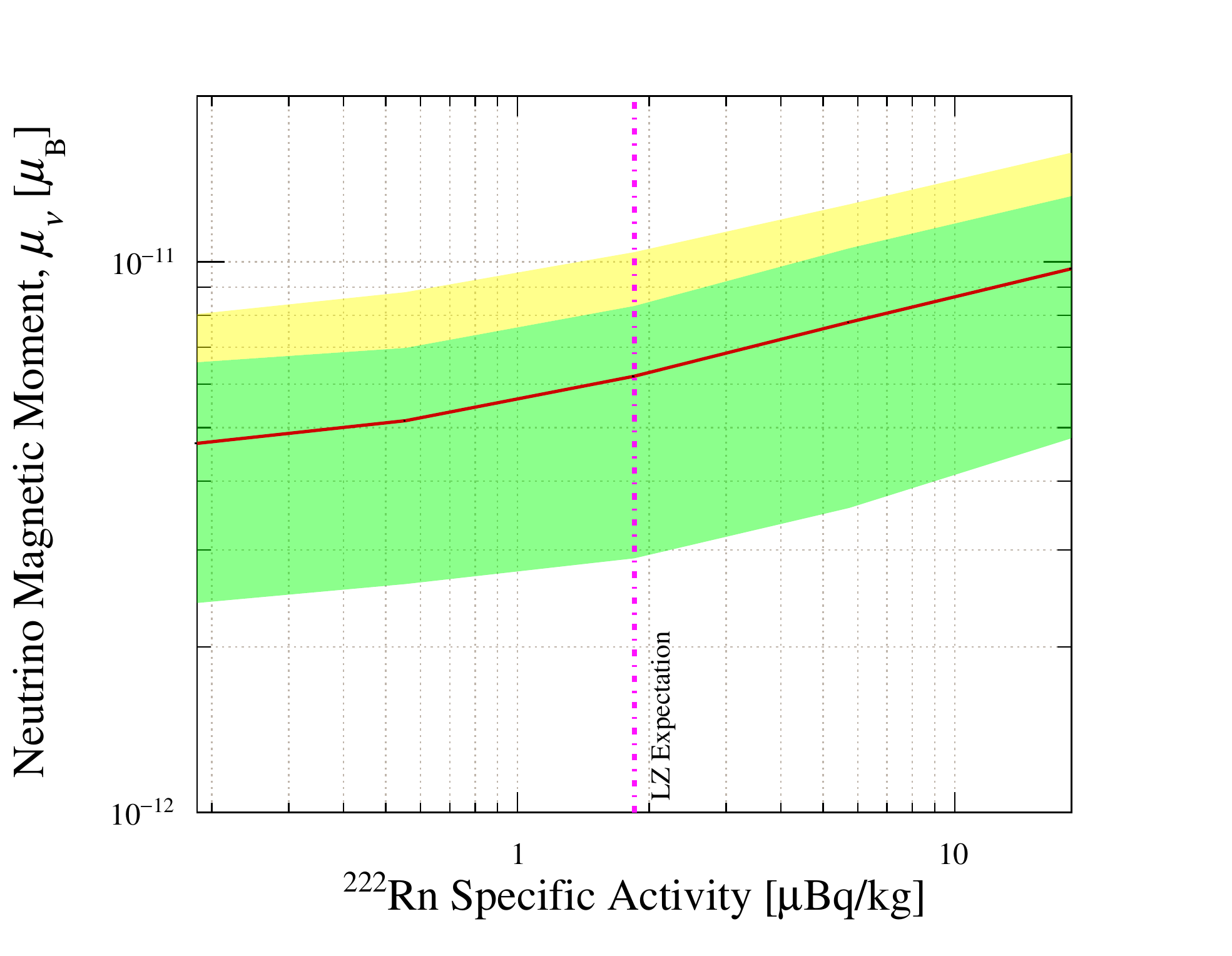}
\caption []{\label{fig:Neu_varyBeta_one}}
\end{subfigure}
\begin{subfigure}{\columnwidth}
\includegraphics[width=\columnwidth]{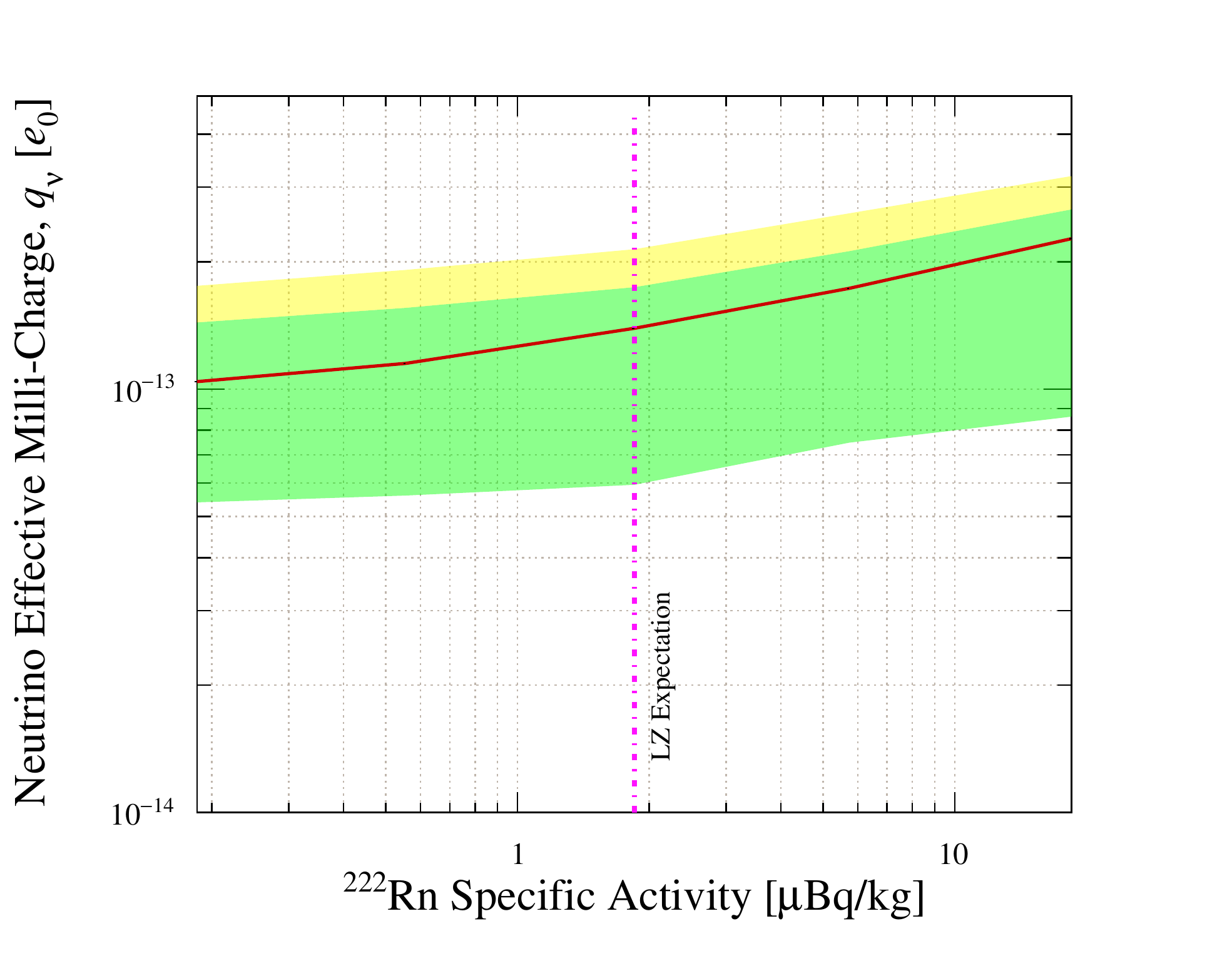}
\caption []{\label{fig:Neu_varyBeta_two}}
\end{subfigure}
\begin{subfigure}{\columnwidth}
\includegraphics[width=\columnwidth]{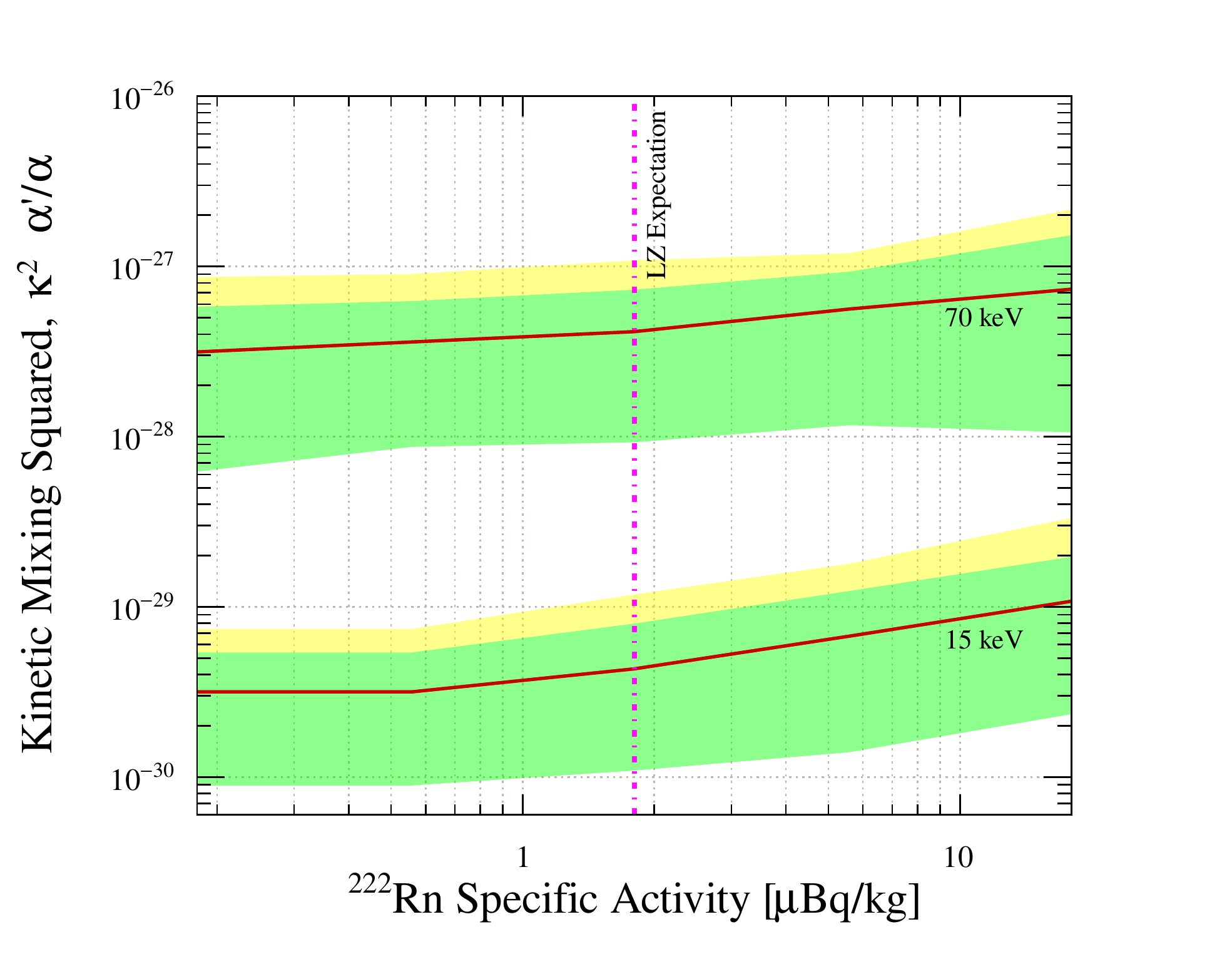}
\caption []{\label{fig:HP_varyBeta}}
\end{subfigure}
\begin{subfigure}{\columnwidth}
\includegraphics[width=\columnwidth]{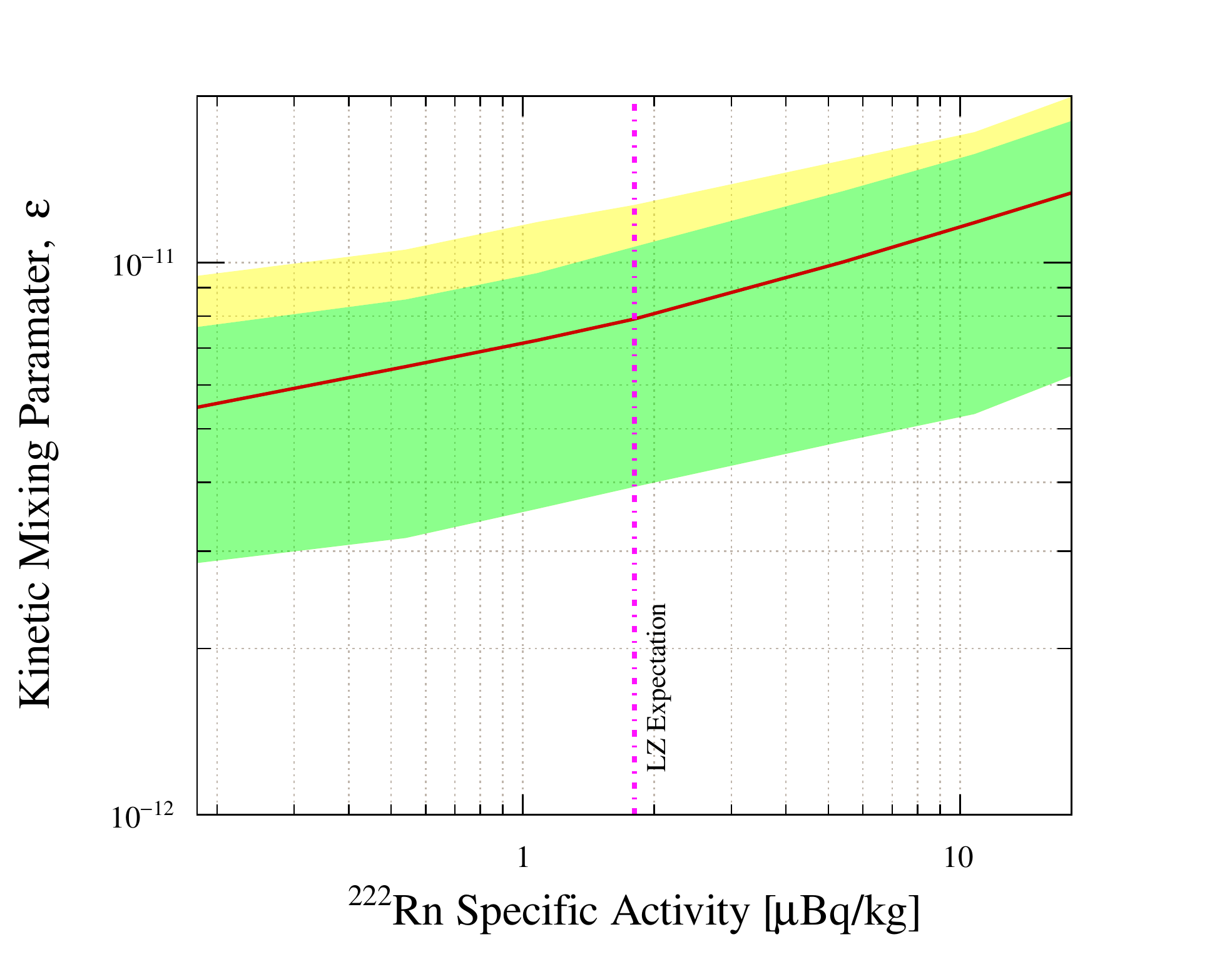}
\caption []{\label{fig:mdm_varyBeta}}
\end{subfigure}
\caption{Variation of $90\%$ C.L. exclusion sensitivity to (\subref{fig:Neu_varyBeta_one}) solar neutrino magnetic moment, (\subref{fig:Neu_varyBeta_two}) solar neutrino effective millicharge, (\subref{fig:HP_varyBeta}) kinetic mixing squared, $\kappa^2$ for 15 and 70~keV$/$c$^2$ hidden photons and (\subref{fig:mdm_varyBeta}) mirror dark matter kinetic mixing for local mirror electron temperature 0.3~keV. $\pm 1\sigma$ (green) and $+2\sigma$ (yellow) bands are also shown.} 
\label{fig:SensVaryBeta}
\end{figure*}

Intrinsic Xe contaminants constitute the dominant ER background in LZ at low energies, largely due to $\beta$ decays from the $^{212}$Pb ($^{214}$Pb) progenies of $^{220}$Rn ($^{222}$Rn). These isotopes are especially troublesome as their decays may result in beta particles with no accompanying radiation, either because the beta decay was directly to the ground state, or the associated gamma rays escape without detection. The rate of this dominant background depends strongly on the success of dust-reducing cleanliness protocols during detector assembly and with the somewhat uncertain radon emanation of components in the cold xenon environment.  It is thus relevant to investigate how signal sensitivity varies under differing radon contamination scenarios.  As in the sensitivity projections using the baseline background model (see Table~\ref{tab:backgrounds}), three backgrounds of uniform spatial distribution and similar (nearly flat) spectral shape are grouped together as a single background component:  $^{222}$Rn + $^{220}$Rn + $^{85}$Kr.  We now vary this grouped `distributed betas' component by a factor of 10 greater and smaller than the baseline expectation (keeping the rate uncertainty nuisance parameter proportional at 24\%) and repeat selected exclusion sensitivity projections under these varied background assumptions.  We show the results of these studies in Figure~\ref{fig:SensVaryBeta}, plotting projected 90\% CL exclusion sensitivity as a function of the dominating $^{222}$Rn portion of the varied background component.

Given the 1000-day exposure time assumed, any possible signal competes with a large number of background counts, typically numbering in the hundreds of events per keV.  In this `high stats' regime, sensitivity to signal counts changes in proportion to the square root of the background counts in the relevant signal region.  The PLR studies under varied background expectation confirm this expectation at low energies when $^{222}$Rn forms the dominant background.  This square root scaling is seen to weaken in two regimes: first, when the $^{222}$Rn is reduced by a significant factor (such that the solar neutrino scattering rate begins to dominate), or second, when the signal model is constrained to higher energies, above $\sim$40~keV, such that the $\nu\nu\beta\beta$ decay of $^{136}$Xe begins to dominate.  The effect of solar neutrino backgrounds can be seen most clearly as a flattening in the left-most portion of Figure~\ref{fig:Neu_varyBeta_two}, and the effect of $^{136}$Xe $\nu\nu\beta\beta$ can be seen as a difference in slope between the 15~keV and 70~keV HP sensitivities.

As mentioned previously, the shapes of the beta decay spectra that dominate the low-energy background have recently been subject to new theoretical calculation.  This results in a suppression of their rates, at energies relevant to this work, by $\sim$19\% for the dominant $^{214}$Pb species, and smaller amounts for the other sub-dominant species $^{212}$Pb and $^{85}$Kr~\cite{scottha_betas}.  The effect of a $\sim$19\% reduction in low-energy background beta decay rate equates to shifting the 'LZ Expectation' line in Figure~\ref{fig:Neu_varyBeta_two} to the left, from 1.8~$\mu$Bq/kg to 1.46~$\mu$Bq/kg.  It can be seen therefore that these refinements in spectral shape change the projected LZ sensitivities of this work by less than 10\%.

The main conclusion from these studies is that LZ sensitivity to these signals will remain world-leading under a variety of reasonable background assumptions.

\section{Conclusion}
\noindent
There are a number of well-motivated extensions to the Standard Model of particle physics that the LZ experiment will be able to test with unprecedented sensitivity. Here we have presented the sensitivity of LZ to theoretical models in which either new particles, such as axions, or new mechanisms of interaction, such as enhanced loop-induced effective electromagnetic neutrino properties, result in additional low-energy ERs. In total seven models are considered, covering a range of signal shape profile and energy.  In each model, LZ is projected to have world-leading sensitivity.  In particular, LZ will thoroughly test any new physics explanation of the recent XENON1T excess.  LZ is currently being commissioned and will commence data-taking at SURF in 2021.

\begin{acknowledgments}
The research supporting this work took place in whole or in part at the Sanford Underground Research Facility (SURF) in Lead, South Dakota. Funding for this work is supported by the U.S. Department of Energy, Office of Science, Office of High Energy Physics under Contract Numbers DE-AC02-05CH11231, DE-SC0020216, DE-SC0012704, DE-SC0010010, DE-AC02-07CH11359, DE-SC0012161, DE-SC0014223, DE-SC0010813, DE-SC0009999, DE-NA0003180, DE-SC0011702,  DESC0010072, DE-SC0015708, DE-SC0006605, DE-SC0008475, DE-FG02-10ER46709, UW PRJ82AJ, DE-SC0013542, DE-AC02-76SF00515, DE-SC0018982, DE-SC0019066, DE-SC0015535, DE-SC0019193 DE-AC52-07NA27344, \& DOE-SC0012447.    This research was also supported by U.S. National Science Foundation (NSF); the U.K. Science \& Technology Facilities Council under award numbers, ST/M003655/1, ST/M003981/1, ST/M003744/1, ST/M003639/1, ST/M003604/1, ST/R003181/1, ST/M003469/1, ST/S000739/1, ST/S000666/1, ST/S000828/1, ST/S000879/1, ST/S000933/1, ST/S000747/1, ST/S000801/1 and ST/R003181/1 (JD); Portuguese Foundation for Science and Technology (FCT) under award numbers PTDC/FIS-­PAR/28567/2017; the Institute for Basic Science, Korea (budget numbers IBS-R016-D1). We acknowledge additional support from the STFC Boulby Underground Laboratory in the U.K., the GridPP~\cite{gridpp1, gridpp2} and IRIS Consortium, in particular at Imperial College London and additional support by the University College London (UCL) Cosmoparticle Initiative. This research used resources of the National Energy Research Scientific Computing Center, a DOE Office of Science User Facility supported by the Office of Science of the U.S. Department of Energy under Contract No. DE-AC02-05CH11231. This work was completed in part with resources provided by the University of Massachusetts' Green High Performance Computing Cluster (GHPCC). The University of Edinburgh is a charitable body, registered in Scotland, with the registration number SC005336. The assistance of SURF and its personnel in providing physical access and general logistical and technical support is acknowledged.  We thank Jiunn-Wei Chen and his group for handing off numerical results from their RRPA calculations, and we thank Patrick Draper for useful discussions.

\end{acknowledgments}

\end{document}